\documentclass[aps,onecolumn,showpacs,superscriptaddress,groupedaddress,amsmath,amssymb]{revtex4-2}  
\usepackage{aas_macros} 
\usepackage{graphicx}   
\usepackage{dcolumn}    
\usepackage{bm}         
\usepackage{makecell}   
\usepackage{mathrsfs}   
\usepackage{color}      
\usepackage[version=4]{mhchem}

\newcommand{\ENDF}{ENDF/B-VIII.0}
\newcommand{\Pu}{$^{239}$Pu}
\newcommand{\CoH}{\texttt{CoH}}
\newcommand{\etal}{\textit{et al.}}
\newcommand{\NJOY}{\texttt{NJOY}}

\graphicspath{{./}{figs/}}

\begin{document}

\hspace{5.2in} \mbox{LA-UR-22-24121}

\title{The Los Alamos evaluation of $^{239}$Pu neutron-induced reactions in the fast energy range}

\author{M.~R.~Mumpower}
\email{mumpower@lanl.gov}
\affiliation{Los Alamos National Laboratory, Los Alamos, NM 87545, USA}

\author{D.~Neudecker}
\affiliation{Los Alamos National Laboratory, Los Alamos, NM 87545, USA}

\author{T.~Kawano}
\affiliation{Los Alamos National Laboratory, Los Alamos, NM 87545, USA}

\author{M.~Herman}
\affiliation{Los Alamos National Laboratory, Los Alamos, NM 87545, USA}

\author{N.~Kleedtke}
\affiliation{Los Alamos National Laboratory, Los Alamos, NM 87545, USA}

\author{A.~E.~Lovell}
\affiliation{Los Alamos National Laboratory, Los Alamos, NM 87545, USA}

\author{I.~Stetcu}
\affiliation{Los Alamos National Laboratory, Los Alamos, NM 87545, USA}

\author{P.~Talou}
\affiliation{Los Alamos National Laboratory, Los Alamos, NM 87545, USA}

\date{October 17, 2022}

\begin{abstract}
A major revision of the evaluation of \Pu{} neutron-induced reaction cross sections is reported in the fast energy range.
The evaluation starts at 2.5 keV incident neutron energy and has been extended up to 30 MeV. 
Several other notable changes are included in this evaluation since the release of \ENDF{} including the adoption of the Standards fission cross section, inclusion of new radiative capture data of Mosby \etal{}, inclusion of the ($n,2n$) data of M\'{e}ot \etal{}, in addition to advances in the treatment of reaction modeling. 
In contrast to previous evaluation efforts, this evaluation is reproducible with detailed information stored chronologically utilizing a \texttt{Git} repository.
The final evaluation results have been compiled into an ENDF-formatted file, which has been processed successfully through \NJOY{}, checked for internal consistency, benchmarked versus older evaluations and validated against a suite of critical assemblies and pulsed-spheres.
\end{abstract}

\maketitle

\section{Introduction}
Accurate and reliable nuclear data are mandated by modern applications. 
The neutron-induced reaction data of \Pu{} have a major impact on nuclear reactor design, fuel-cycle analysis and stockpile stewardship, to name but a few \cite{Talou2011, Ullamnn2013, Becker2013}. 

Recently, new experimental data as well as theoretical model advances have become available that justify revisiting the current \ENDF{} evaluation~\cite{Brown2018}. First of all, new fission cross sections were provided by the Standard committee under the auspices of the IAEA ~\cite{Neudecker2020, Neudecker2021}, and in part motivated by the release of fissionTPC data~\cite{Snyder2019}. A new evaluation of the Prompt Fission Neutron Spectrum was performed recently~\cite{Neudecker2022PFNS}, making use of recent Chi-Nu and CEA data taken at LANSCE~\cite{Kelly2021,Marini2020,NNSACEA} that cover for the first a broad incident and outgoing neutron energy range with high precision. 
Also, a new, high-precision average prompt fission neutron multiplicity measurement by the CEA/ NNSA collaboration~\cite{Marini2021} was included, together with modeling from the CGMF fission event generator~\cite{Talou2021}, leveraging a new evaluation by LANL collaborators (Neudecker, Lovell, Talou)~\cite{Neudecker2021nu}.
A new measurement of the capture cross sections was performed at LANL by Mosby \etal~\cite{Mosby2018} and additional experimental results became available for the ($n,2n$) reaction by M\'{e}ot~\cite{Meot2021}. 
The resolved resonance region (RRR) and unresolved resonance region (URR) parameters have been updated by M. Pigni \etal{} \cite{Pigni2022} since \ENDF{}. 
These important changes are combined with advances in nuclear reaction theory modeling in the fast energy range such as including M1 scissors mode in the gamma emission, Engelbrecht-Weidenmueller transformation accounting for the interference between direct and compound nucleus mechanisms. 
The availability of new data and upgraded theoretical efforts afford essential improvements to the evaluation of \Pu{} cross sections. 
 
The new cross sections in the fast neutron range were calculated using modern reaction code \CoH{}~\cite{Kawano2019, Kawano2021a, Kawano2021b} with input parameters carefully adjusted to reproduce experimental data. 
The whole procedure was encapsulated in Python scripts and is perfectly reproducible. 
This is essential for extending this evaluation procedure to other Pu isotopes using a consistent set of parameters when experimental data are scarce. 

The impact of those expected improvements in the fundamental nuclear data stored in evaluated libraries has been carefully studied against a suite of critical assemblies as well as LLNL pulsed-spheres, which are reviewed at the end of this document.

\section{Evaluation Methodology}
We evaluate neutron induced cross sections of \Pu{} by combining together differential experimental datasets with theoretical modeling. 
Reaction cross section data are obtained primarily from the EXFOR repository of experimental data \cite{Otuka2014}. 
More recent individual datasets not yet available in EXFOR are also included on a case-by-case basis. 
We tabulate the individual datasets used for the evaluation of each reaction channel in the sections that follow. 

To model cross sections, we employ the Los Alamos statistical Hauser-Feshbach code, \CoH{}~\cite{Kawano2019, Kawano2021a, Kawano2021b}. 
Nuclear structure properties are included from the RIPL-3 database \cite{Capote2009}. 
This code supports an extension of the Hauser-Feshbach theory to the coupled-channels scheme for deformed nuclei, as is the case with the target system, \Pu{}. 
Direct and semi-direct reaction mechanisms, as well as the pre-equilibrium process are also supported in this code. 

A Bayesian approach is taken to fit model cross sections to experimental data following the generalized techniques outlined in Hobson \textit{et al.} (2002) \cite{Hobson2002}. 
Individual datasets, enumerated by the variable $k$, are assumed to be independent. 
Overall the total fit to data is minimized by considered the model discrepancy with each dataset,
\begin{equation}
    \label{eqn:chi2}
    \chi^2_k = (\pmb{D}_k - \pmb{\mu}_k)^T \pmb{V}^{-1}_k (\pmb{D}_k - \pmb{\mu}_k) \ ,
\end{equation}
where $\pmb{D}_k$ is a measured cross section, $\pmb{\mu}_k$ is its expected value from a model and $\pmb{V}_k$ is the covariance matrix. 
The likelihood for each dataset is defined as 
\begin{equation}
    \label{eqn:likelihood}
    \textrm{Pr}(\pmb{D} | \theta, \alpha_k) = \frac{1}{\pi^{n_k/2}|\pmb{V}_k|^{1/2}}(\alpha^{n_k/2}_k)\textrm{exp}(-\frac{1}{2} \alpha_k \chi^2_k) \ .
\end{equation}
The $\alpha_k$ represent a weighting for how important a particular individual dataset is to the fit. 
It is the choice of the evaluator how to handle $\alpha_k$. 
Some datasets are very important in which case they make take on large values, while a relatively inconsequential dataset exhibit smaller values of $\alpha_k$. 
A value of $\alpha_k=0$ represents a dataset that is not included in the fit. 

The total likelihood of independent datasets is a product,
\begin{equation}
    \label{eqn:total_likelihood}
    \textrm{Pr}(\pmb{D} | \theta) = \prod_{k=1}^{N} \frac{2 \Gamma(\frac{n_{k}}{2} + 1)}{\pi^{n_k/2}|\pmb{V}_k|^{1/2}}(\chi^2_k+2)^{-(\frac{n_k}{2}+1)} \ ,
\end{equation}
where $\theta$ is the set of model parameters, $n_k$ denotes the number of measured data points in the $k$-th dataset, and $\Gamma$ is the mathematical function defined as $\Gamma \left( x \right) \equiv \int\limits_0^\infty {s^{x - 1} e^{ - s} ds}$. 
Bayes' theorem is invoked to find the posterior distribution,
\begin{equation}
    \label{eqn:posterior}
    \overline{\textrm{Pr}}(\theta | \pmb{D}) = \textrm{Pr}(\pmb{D} | \theta)\textrm{Pr}(\theta)
\end{equation}
where $\textrm{Pr}(\pmb{D} | \theta)$ is the aforementioned total likelihood and $\textrm{Pr}(\theta)$ is the prior distribution. 
In this derivation, it is assumed that experimental data uncertainties follows a normal distribution. 
Using this methodology, there is a strict hierarchy of importance: experimental data is valued above all else, followed by model, followed by derived data that implicitly use model(s) to ascertain a value. 

The procedure outlined above is computationally demanding, where Monte Carlo techniques are generally invoked. 
In order to maintain model consistency throughout the evaluation, we use an incremental approach. 
We first begin by fitting the model to available total cross section data. 
This procedure fixes the associated optical model parameters and the resulting neutron transmission coefficients used in the Hauser-Feshbach model. 
Next, we set fission parameters to closely match the Standards fission cross section of Neudecker \etal~\cite{Neudecker2020, Neudecker2021}. 
Although the final evaluated fission cross section present in our ENDF-formatted file is taken from the Standards, it is important to calculate this channel relatively well as it impacts all other competing channels computed in \CoH. 
We then study and fix subsequent parameters of the remaining channels in a procedural fashion. 
When the model is not sufficient to describe differentiate data, as in the case of the (n,$\gamma$) channel, the model may be used as a Bayesian prior in the above framework to generate an evaluated cross section. 
Model parameters are optimized in the fit so that they may serve as a basis for future evaluation efforts along the Pu isotopic chain where data is sparse. 

Our evaluation effort is recorded incrementally using \texttt{Git} version control. 
Each decision is encapsulated in a single `commit' of the repository and reflects a major choice/change in the evaluation. 
As an example, a single commit may be an update to model parameters. 
Another commit may contain information relevant to the inclusion or exclusion of particular data. 
The different codes of the evaluation are tied together using version 3 of the Los Alamos \texttt{NEXUS} computational nuclear data platform \cite{Mumpower2020}. 
\texttt{NEXUS} is written in the Python programming language and provides the interface between codes and data. 
With this methodology it is possible to transfer, run and reproduce the evaluation on any modern computer. 

\section{Total}

The target nucleus, \Pu{}, is deformed in its ground state and has therefore to be treated in the coupled-channel formalism \cite{Kawano2021a}. 
The total cross section is calculated with the \CoH{} code with coupling between the ground state and the first six excited states ($E_\textrm{x} \sim 7.861$ keV through $E_\textrm{x} \sim 318.5$ keV). 
These values are listed in Table \ref{tab:cc_levels}. 

\def\arraystretch{1.5}%
\begin{table}
\centering
\caption{Table of energy levels of \Pu{} used in the coupled channels calculation. }
\label{tab:cc_levels}
\begin{tabular}{|c|c|c|} 
\hline
\textbf{Level number} & \textbf{Excitation Energy (MeV)} & \textbf{Spin-Parity (J$^{\pi}$)} \\ \hline
1 & 0.000000 & $\frac{1}{2}^{+}$  \\ \hline
2 & 0.007861 & $\frac{3}{2}^{+}$  \\ \hline
3 & 0.057275 & $\frac{5}{2}^{+}$  \\ \hline
4 & 0.075705 & $\frac{7}{2}^{+}$  \\ \hline
5 & 0.163760 & $\frac{9}{2}^{+}$  \\ \hline
6 & 0.192800 & $\frac{11}{2}^{+}$ \\ \hline
7 & 0.318500 & $\frac{13}{2}^{+}$ \\ \hline
\end{tabular}
\end{table}

The static quadrupole deformation is set to $\beta_2 = 0.202$, slightly lower than the FRDM-calculated value of 0.236~\cite{Moller2016}. 
This value remains fixed during the optimization of the coupled channels calculation. 
The Soukhovitskii potential is used as the basis for the optical model \cite{Soukhovitskii2005, Capote2005}. 
The total cross section is the first channel to be fit in our hierarchical modeling procedure. 

To optimize the calculated total cross section with respect to available data, $\sigma_\textrm{T}$, a Metropolis random-walk algorithm is used to probe the optical model parameter space. 
The parameter space consists of six parameters, including the potential depths, diffuseness and radii. 
All other model parameters are held fixed during this optimization. 
It was determined that holding the optical model deformation fixed, rather than letting it vary during the optimization was ideal for approximating the global minimum. 

The optimal fit is found by looping over the selected experimental datasets listed in Table~\ref{tab:total} and minimizing the chi-square goodness of fit via Bayesian hyperparameter optimization, see e.g.~Ref.~\cite{Hobson2002}. 
The weight of each dataset to the overall fit is approximately determined by the inverse of its mean reported uncertainty in the energy region. 
The previous \ENDF{} evaluation of the total cross section is also included in the fit and given the highest relative weighting. 

\begin{figure}
 \begin{center}
 \includegraphics[width=\textwidth]{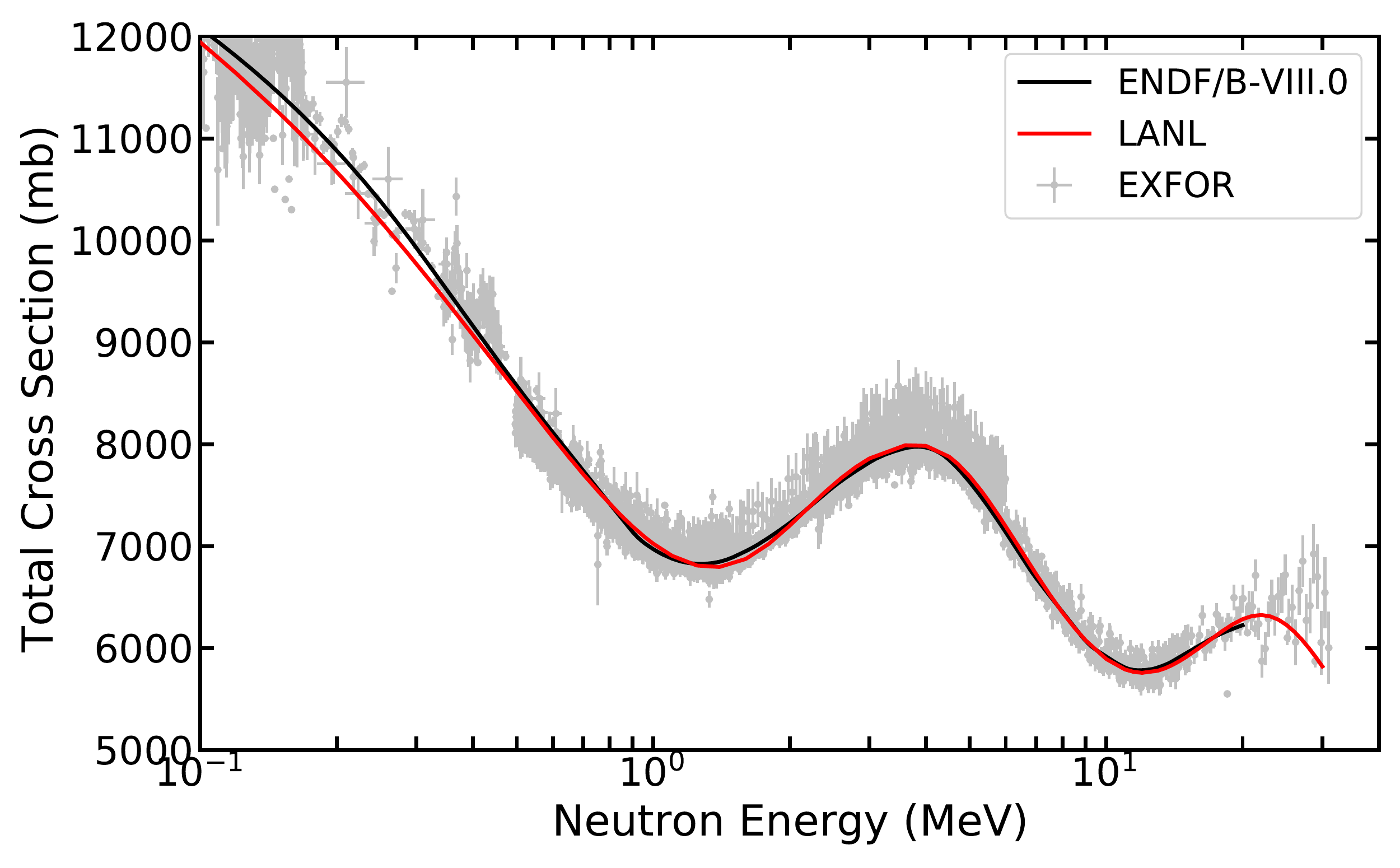}
 \caption{The total cross section above 100 keV. In contrast to \ENDF{} (black line), the current evaluation extends to an incident neutron energy of 30 MeV. }
 \label{fig:cs_total}
 \end{center}
\end{figure}

This procedure leads to the total cross section fit as shown in Figure \ref{fig:cs_total}. 
An excellent reproduction of the previous \ENDF{} evaluation is observed. 
Slight modifications are seen relative to \ENDF{} below 30 keV where the reported uncertainties of the datasets pull down the fit to the total cross section. 
A similar modification, albeit to a much smaller effect, can be found between 1 and 2 MeV near the local minimum of $\sigma_\textrm{T}$. 

\def\arraystretch{1.5}%
\begin{table}
\centering
\caption{Experimental datasets and previous evaluations used in the evaluation of the neutron-induced total cross section of \Pu{} that have at least one measurement in the fast energy range. Entries in the table are sorted by increasing year of publication. }
\label{tab:total}
\begin{tabular}{|c|c|c|c|c|} 
\hline
\textbf{First Author} & \textbf{Energy Range (MeV)} & \textbf{Year} & \textbf{Reference} & \textbf{EXFOR \# } \\ \hline

R.L.~Henkel    & 0.041-20.5          & 1952 & \cite{Henkel1952} & 12396005 \\ \hline
J.H.~Coon      & 14.12-14.12         & 1952 & \cite{Coon1952} & 12524002 \\ \hline
C.T.~Hibdon    & 0.00115-0.159       & 1954 & \cite{Hibdon1954} & 11002006 \\ \hline
P.A.~Egelstaff & 2.25$10^{-7}$-0.07933    & 1958 & \cite{Egelstaff1958} & 21032002 \\ \hline
A.~Bratenahl   & 7.05-14.25          & 1958 & \cite{Bratenahl1958} & 11155035 \\ \hline
J.M.~Peterson  & 17.3-28.9           & 1960 & \cite{Peterson1960} & 11108041 \\ \hline
D.~G.Foster    & 2.258-14.965        & 1971 & \cite{Foster1971} & 10047100 \\ \hline
J.~Cabe        & 1.105-1.07          & 1973 & \cite{Cabe1973} & 20480022 - 20480023 \\ \hline
K.A.~Nadolny   & 0.5-30.984          & 1973 & \cite{Nadolny1973} & 10589002 \\ \hline
A.B.~Smith     & 0.65055-1.499       & 1973 & \cite{Smith1973} & 10212002 \\ \hline
R.B.~Schwartz  & 0.496-15.17         & 1974 & \cite{Schwartz1974} & 10280007 \\ \hline
W.P.~Poenitz   & 0.048-4.807         & 1981 & \cite{Poenitz1981} & 10935008 \\ \hline
W.P.~Poenitz   & 1.818-20.91         & 1983 & \cite{Poenitz1983} & 12853056 \\ \hline
R.R.~Spencer   & 3.2977$10^{-7}$-0.016387 & 1987 & \cite{Spencer1987} & 12941006 \\ \hline
J.A.~Harvey    & 2.8049$10^{-5}$-0.01553  & 1988 & \cite{Harvey1988} & 13632002 - 13768011 \\ \hline
\end{tabular}
\end{table}

\section{Fission}
The neutron-induced fission cross section is taken from the Standard evaluation work of D. Neudecker \etal~\cite{Neudecker2020, Neudecker2021}. 
To avoid repetition, we refer the reader to these external documents for appropriate details. 
Here, we discuss the parameters used in our modeling that closely reproduce the total fission cross section.

The transmission coefficient for fission is approximated using the Hill-Wheeler formula assuming transmission through a parabolic barrier \cite{Hill1953},
\begin{equation}
 T_{\textrm f}(E) = \frac{1.0}{1.0+\exp(2\pi \frac{B_{\textrm f}-E}{C})} \ ,
 \label{eqn:Tf}
\end{equation}
where $B_{\textrm{f}}$ is the fission barrier height, $C$ is the associated curvature and $E$ is the relative excitation energy of the compound nucleus. 
The fission barrier parameters (height and curvature) set the form of the inverted parabolic barrier. 
These parameters are listed in the Table \ref{tab:nfb_params}. 
The effective transmission coefficient for fission depends on the number of fission barriers adopted for the particular nucleus. 
In this work we use a double humped barrier for each nucleus and thus the effective fission transmission coefficient is,
\begin{equation}
 T^{eff}_{\textrm f} = \frac{T_{\textrm A} T_{\textrm B}}{T_{\textrm A} + T_{\textrm B}} \ ,
 \label{eqn:Teff}
\end{equation}
where $T_{\textrm A}$ and $T_{\textrm B}$ are the first and second fission transmission coefficients for each barrier respectively. 

\def\arraystretch{1.5}%
\begin{table}
\centering
\caption{The parabolic fission barrier parameters used in the \CoH{} statistical Hauser-Feshbach code. Note that for each nucleus along the isotopic chain, a double humped barrier is used consisting of an inner and outer barrier.}
\label{tab:nfb_params}
\begin{tabular}{|c|c|c|c|} 
\hline
\textbf{Nucleus} & \textbf{Inner / outer barrier} & \textbf{Barrier height (MeV)} & \textbf{Curvature ($\hbar \omega$)}  \\ \hline
$^{240}$Pu & inner & 5.76 & 0.44 \\ \hline
$^{240}$Pu & outer & 5.03 & 0.77 \\ \hline

$^{239}$Pu & inner & 5.75 & 0.52 \\ \hline
$^{239}$Pu & outer & 5.45 & 0.68 \\ \hline

$^{238}$Pu & inner & 5.55 & 0.60 \\ \hline
$^{238}$Pu & outer & 5.20 & 0.80 \\ \hline

$^{237}$Pu & inner & 5.80 & 0.80 \\ \hline
$^{237}$Pu & outer & 5.80 & 0.52 \\ \hline

$^{236}$Pu & inner & 6.00 & 1.04 \\ \hline
$^{236}$Pu & outer & 5.00 & 0.60 \\ \hline

\end{tabular}
\end{table}

\begin{figure}
 \begin{center}
 \includegraphics[width=\textwidth]{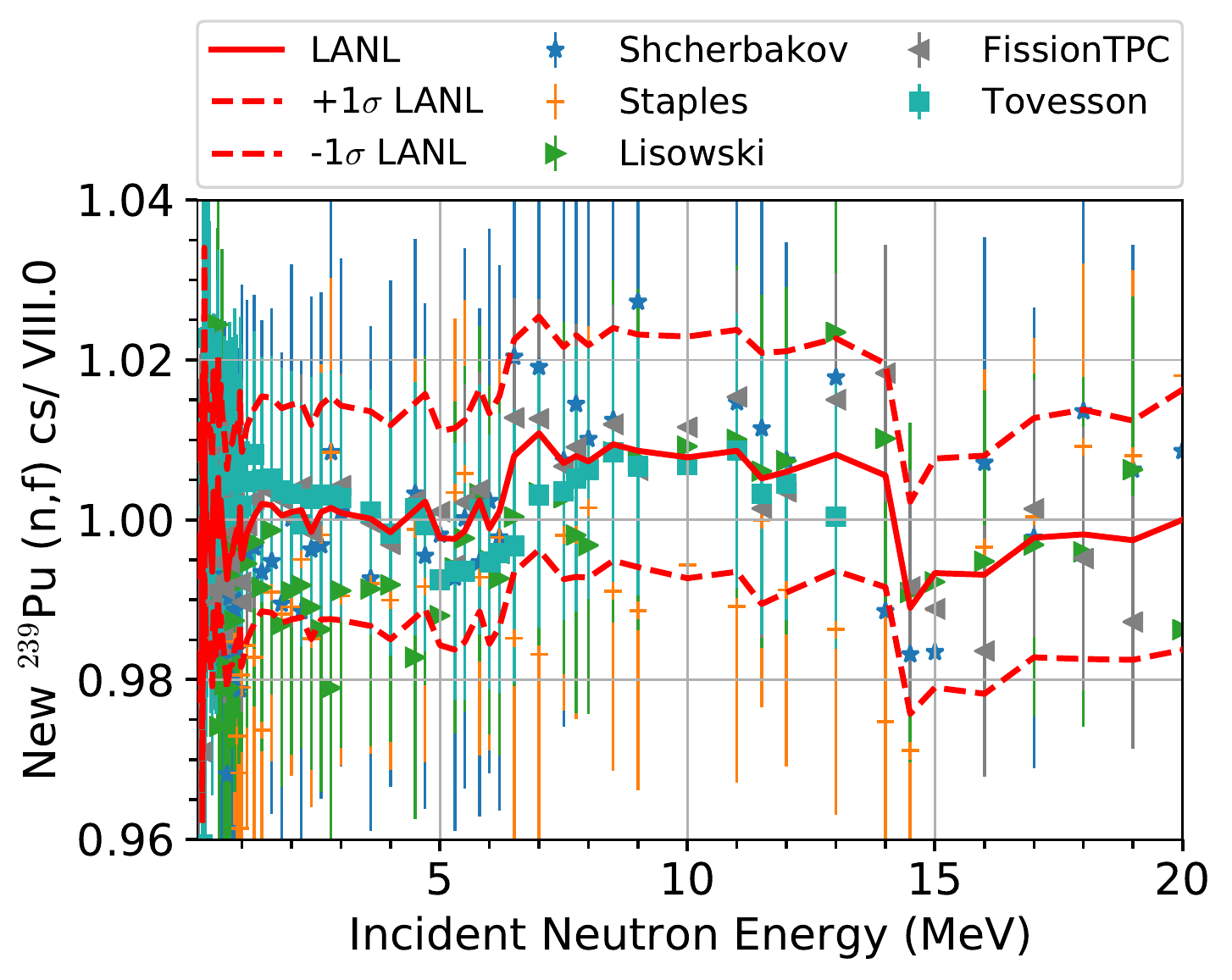}
 \caption{The ratio of the evaluated LANL fission cross section with respect to \ENDF{}. Selected experimental data are shown that raised questions for the standards 2018 release. Newest FissionTPC data are also shown.}
 \label{fig:cs_fis_ratio}
 \end{center}
\end{figure}

\begin{figure}
 \begin{center}
 \includegraphics[width=\textwidth]{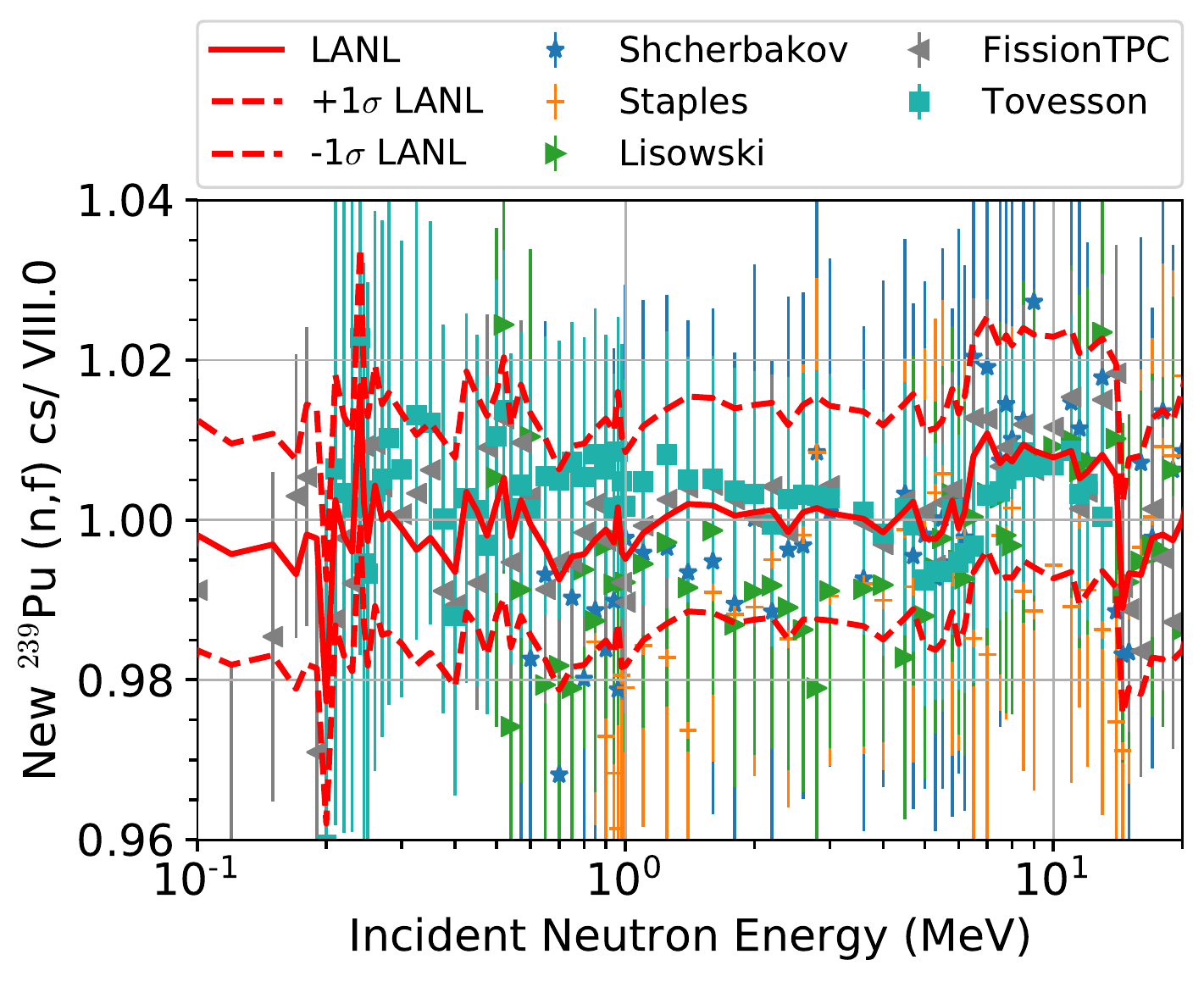}
 \caption{The ratio of the evaluated LANL fission cross section with respect to \ENDF{} on a log scale. Selected experimental data are shown that raised questions for the standards 2018 release. Newest FissionTPC data are also shown.}
 \label{fig:cs_fis_ratio_log}
 \end{center}
\end{figure}

The ratio of the fission cross section with respect to \ENDF{} is shown in Figure~\ref{fig:cs_fis_ratio} across the entire incident energy range considered in \ENDF{}. 
The effect of the FissionTPC and Tovesson data has been to increase the fission cross section, particularly between 7 and 14 MeV. 
The lower incident energy dependence can be observed by using a log scale on the X-axis as in Figure~\ref{fig:cs_fis_ratio_log}. 

\section{Elastic \& Inelastic}

A recent theoretical improvement in the treatment of the direct reaction channels in a coupled-channel calculation is the diagonalization of the S-matrix using the Engelbrecht-Weidenm\"{u}ller (EW) transformation \cite{Engelbrecht1973}. This development was implemented~\cite{Kawano2016} in the \CoH\ code, resulting in a more realistic treatment of the inelastic cross sections when the direct channels are strongly coupled, which is the case in the study of low-energy neutron-induced reactions on $^{239}$Pu.

\begin{figure}
 \begin{center}
 \includegraphics[width=\textwidth]{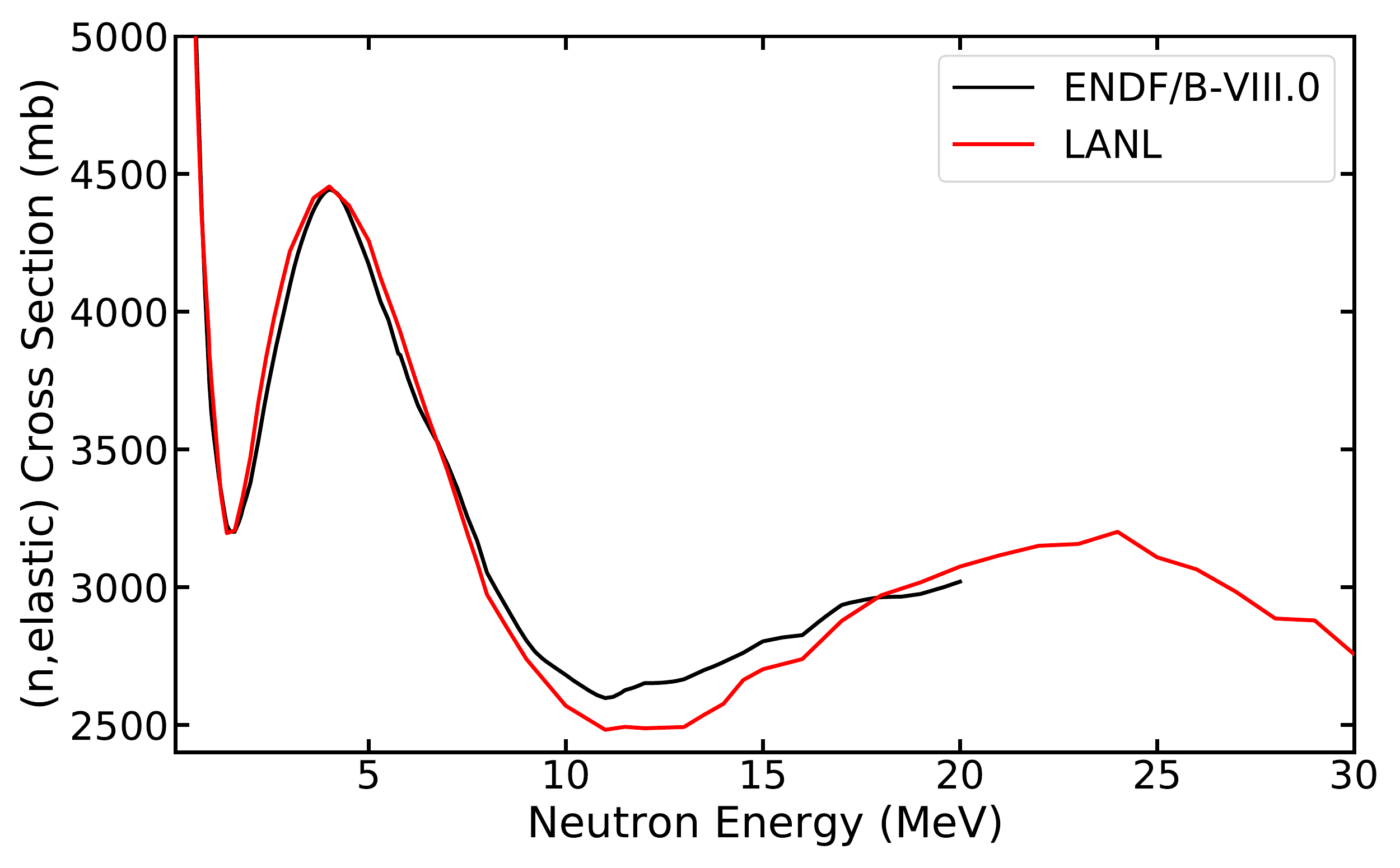}
 \caption{The elastic cross section over the entire evaluated along with comparison to \ENDF{}.}
 \label{fig:cs_el}
 \end{center}
\end{figure}

The elastic cross section for \Pu{} is shown in Figure \ref{fig:cs_el}. 
In general, our evaluation closely follows \ENDF{}. 
Between $\sim$8 and $\sim$18 MeV incident energy, the cross section is slightly lower than \ENDF{} owing to a relatively stronger inelastic channel. 

\begin{figure}
 \begin{center}
 \includegraphics[width=\textwidth]{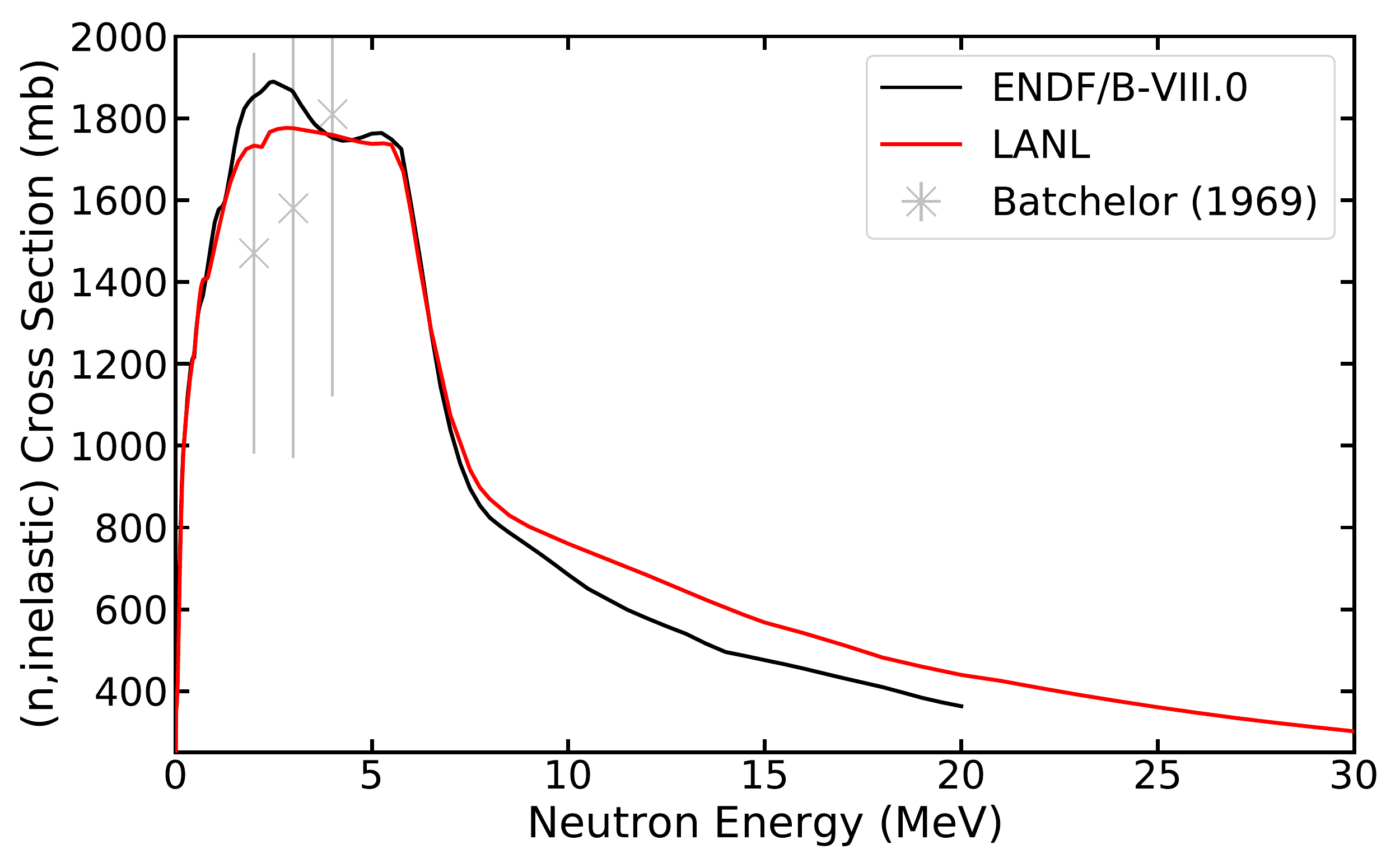}
 \caption{The inelastic cross section over the entire evaluated range along with comparison to \ENDF{}.}
 \label{fig:cs_inel}
 \end{center}
\end{figure}

\begin{figure}
 \begin{center}
 \includegraphics[width=\textwidth]{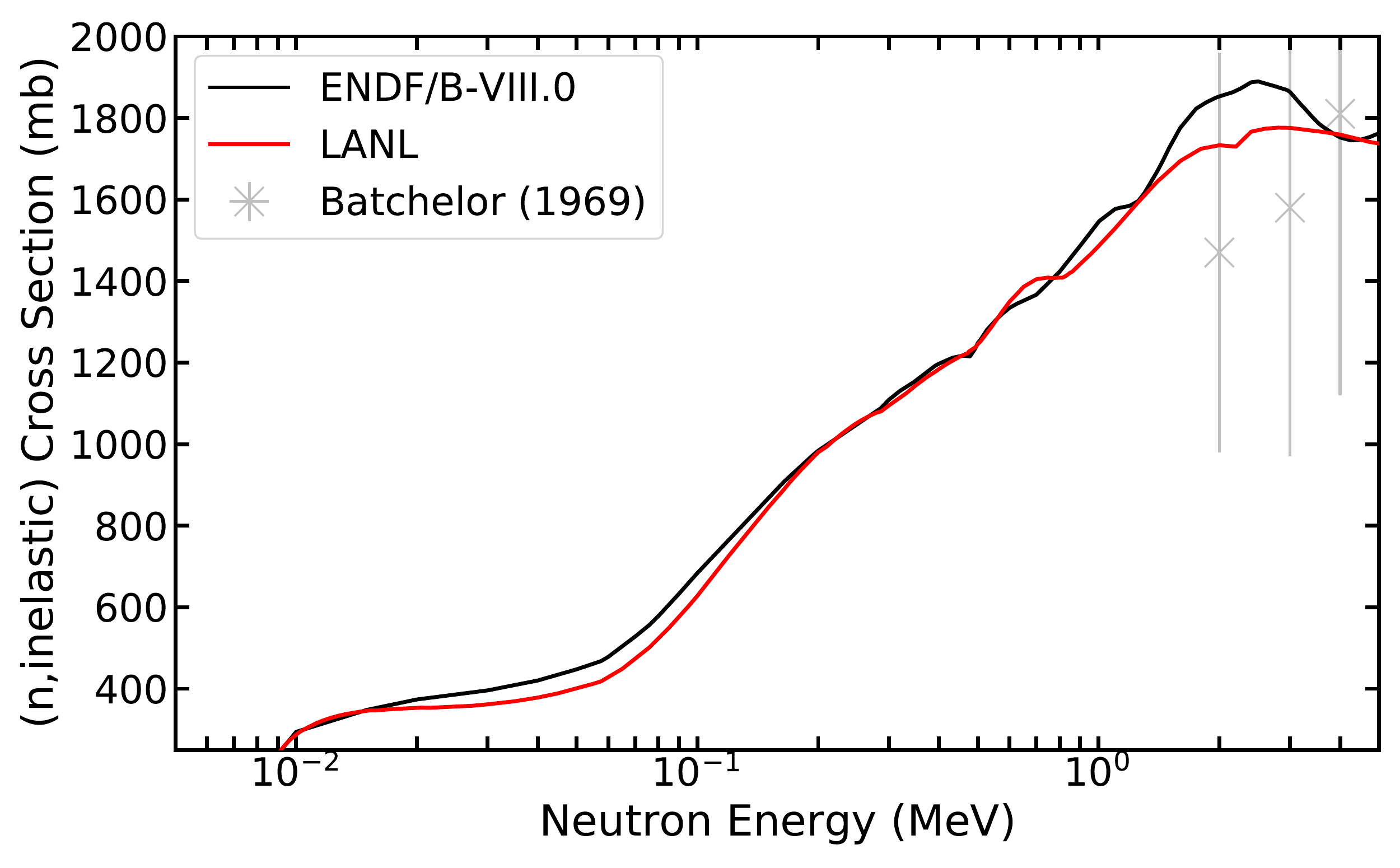}
 \caption{The inelastic cross section near threshold along with comparison to \ENDF{}.}
 \label{fig:cs_inel_close}
 \end{center}
\end{figure}

The inelastic cross section is shown in Figure \ref{fig:cs_inel} over the entire energy range. 
Good agreement is found between our evaluation and \ENDF{} with the most notable deviations arising at the missing ENDF hump near 2 MeV and our cross section being larger above 10 MeV. 
The first deviation can be seen better in Figure \ref{fig:cs_inel_close}, which highlights the behavior of the evaluation near threshold. 

It is well known that deformed nuclei at relatively low excitation energies exhibit a collective nature. 
This observation can be modeled by the Quasi-particle Random Phase Approximation (QRPA), as shown by Kerveno \etal{} \cite{Kerveno2021}. 
The collective excitation can be interpreted as an effective enhancement in the partial state density for 1-particle-1-hole configurations. 
We therefore update the base \CoH{} model calculation to include a modification to the pre-equilibrium level density. 
Specifically, the collective enhancement is added to the 1-particle-1-hole state density in the exciton model. 
The strength of this enhancement is governed by a single parameter which is fit to LLNL pulsed-sphere neutron-leakage spectra \cite{Wong1972}. 

\begin{figure}
 \begin{center}
 \includegraphics[width=\textwidth]{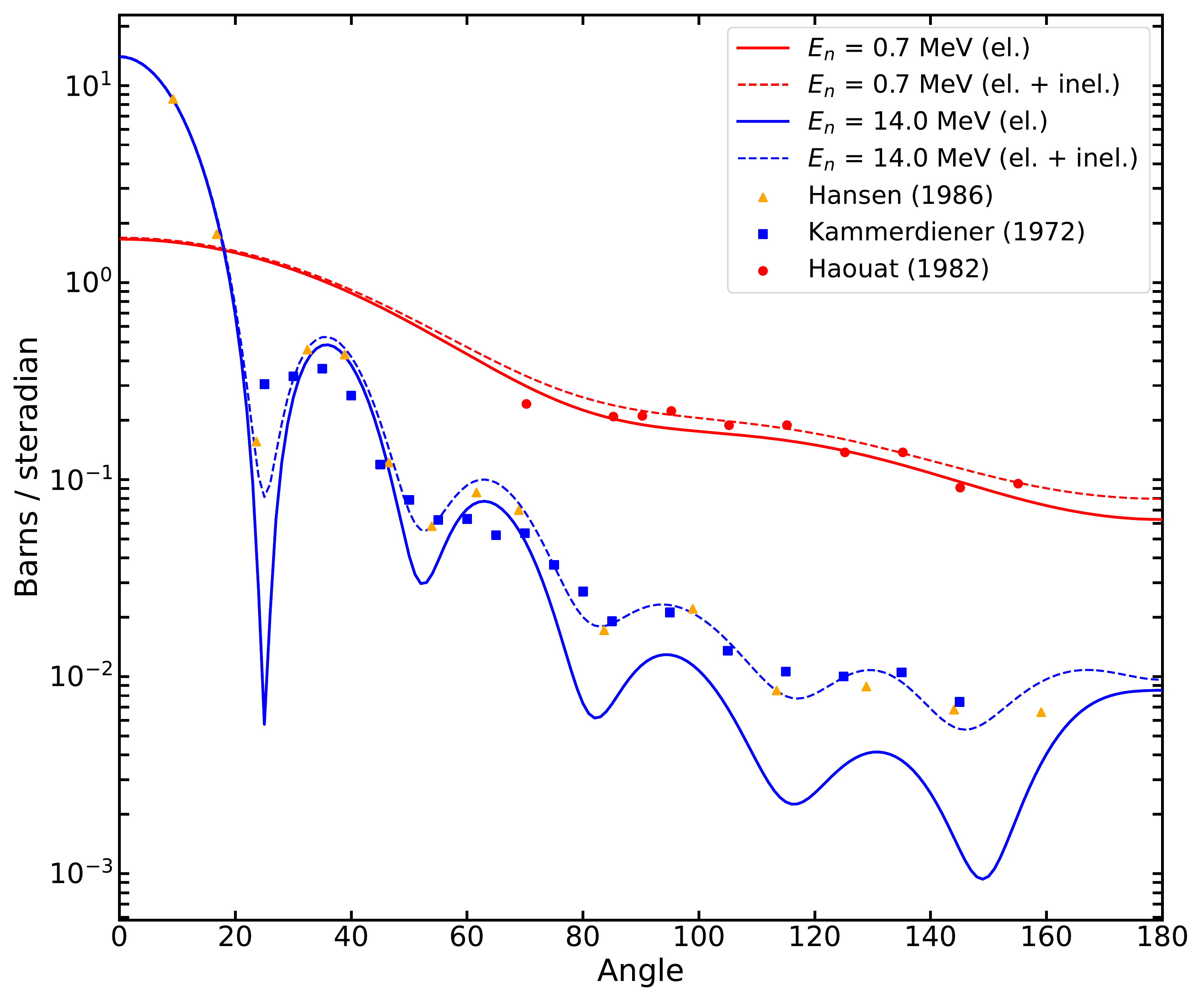}
 \caption{Angular distribution at 0.7 and 14.0 MeV incident neutron energies. The elastic component is shown by a solid line while the dashed lines are the sum of elastic and inelastic. }
 \label{fig:cs_angdist_E0.7}
 \end{center}
\end{figure}

Comparisons are made with select datasets to showcase the the evaluated elastic and inelastic cross sections. 
Figure \ref{fig:cs_angdist_E0.7} shows the angular distributions at incident neutron energies of $E_n=0.7$ and $E_n = 14$ MeV. 
Note that inclusion of the first inelastic level, as shown by the dashed curves, greatly improves the match to the measured data of Haouat \textit{et al.} (1982) \cite{Haouat1982} at 0.7 MeV and Hansen \etal\ (1986) \cite{Hansen1986} and Kammerdiener \etal\ (1972) \cite{Kammerdiener1972} at 14 MeV. 

\begin{figure}
 \begin{center}
 \includegraphics[width=\textwidth]{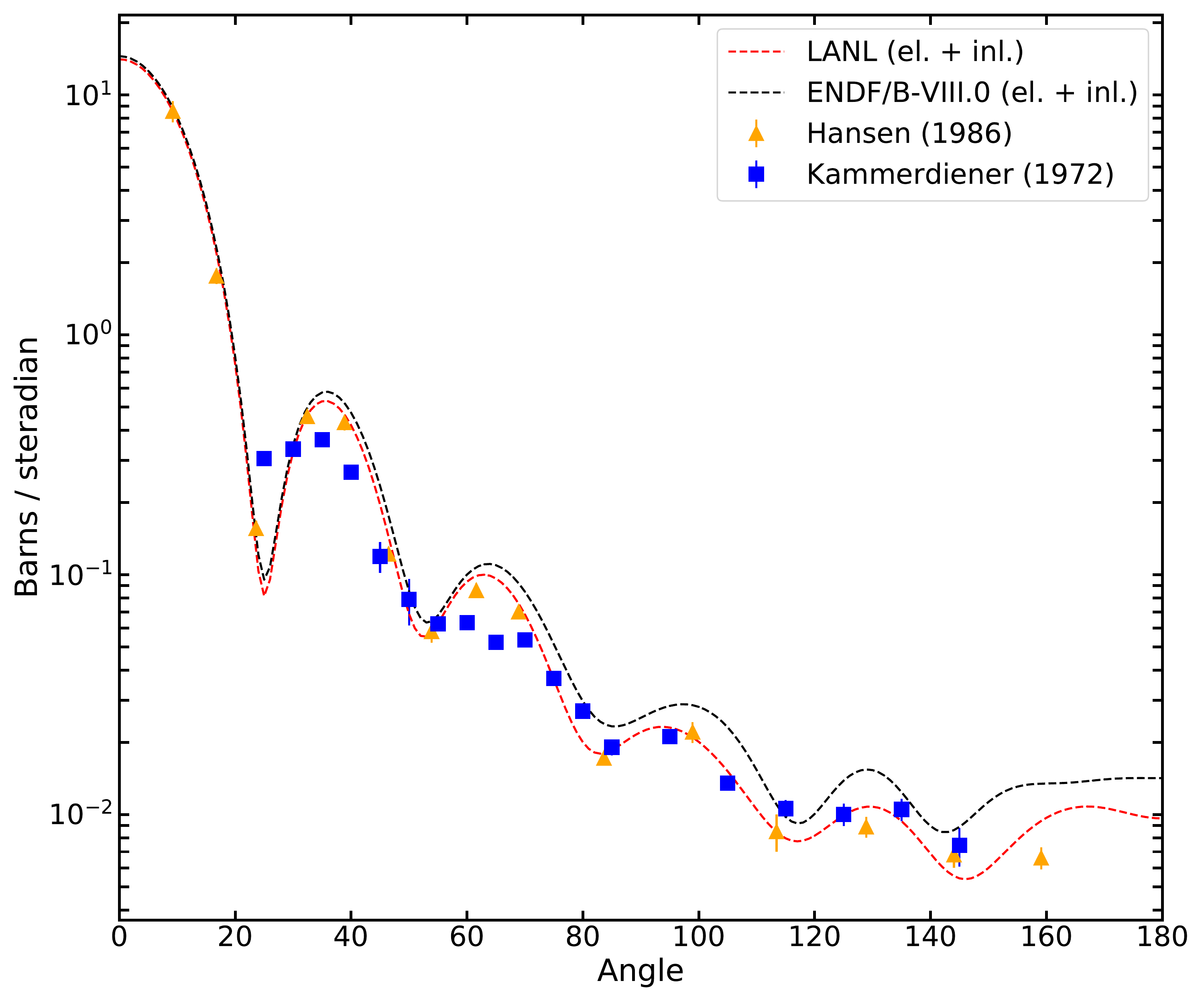}
 \caption{The angular distribution at 14 MeV incident neutron energy as compared with \ENDF{}. }
 \label{fig:cs_angdist_E14}
 \end{center}
\end{figure}

The angular distribution at 14 MeV is plotted again in Figure~\ref{fig:cs_angdist_E14}, this time comparing with \ENDF{}. 
Here there is clear improvement over \ENDF{}, which generally overshoots data.  

\begin{figure}
 \begin{center}
 \includegraphics[width=\textwidth]{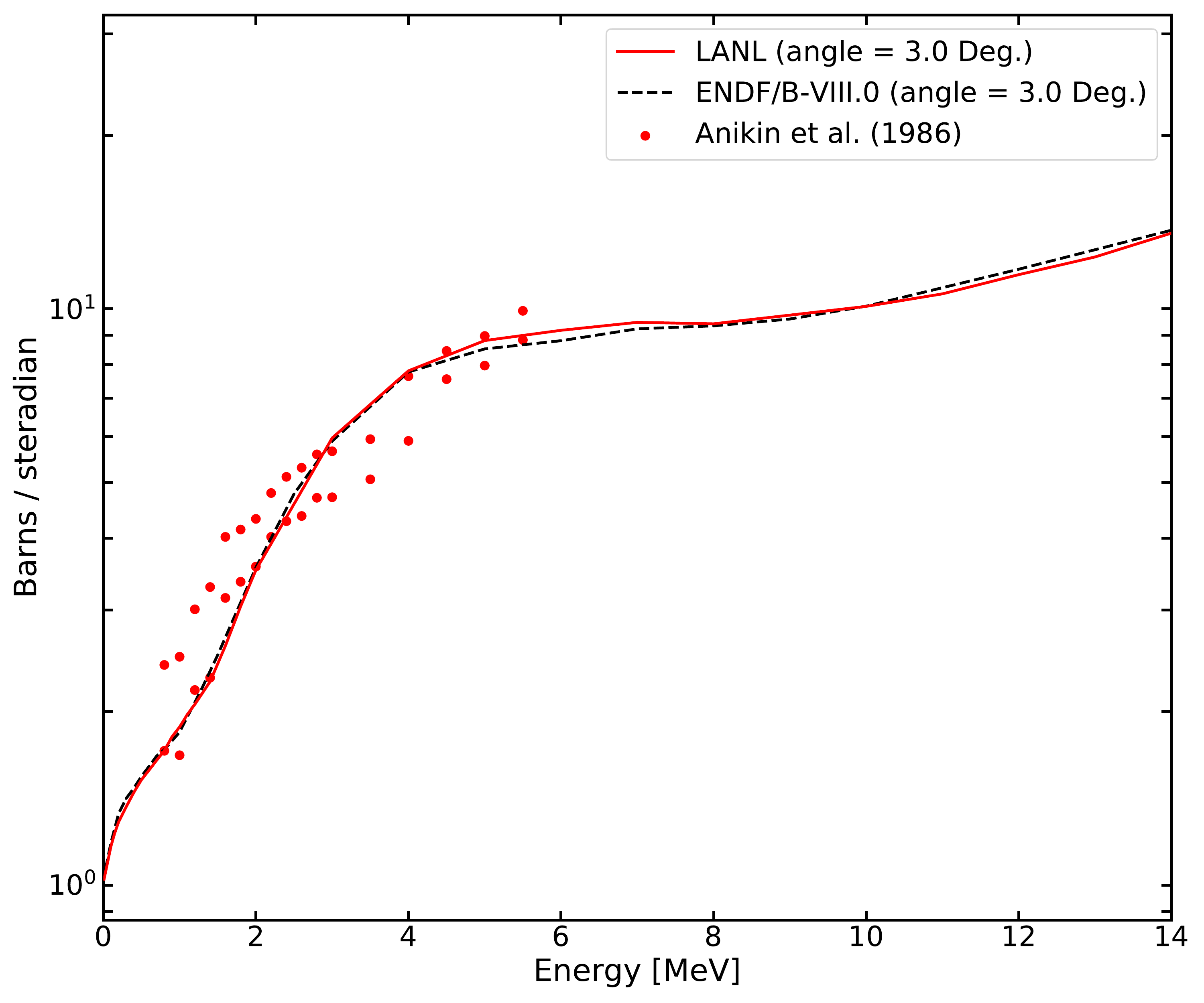}
 \caption{Elastic angular distribution at fixed forward angle as a function of energy.}
 \label{fig:cs_el_angdist2}
 \end{center}
\end{figure}

The elastic angular distribution at fixed forward angle of 3 degrees is shown in Figure \ref{fig:cs_el_angdist2}. 
Overall the new evaluation and \ENDF{} perform similarly. 
A slight shift upwards relative to \ENDF{} is noted around 5 MeV which brings the evaluation in closer alignment with the data of Anikin \textit{et al.} (1986) \cite{Anikin1986}. 

\begin{figure}
 \begin{center}
 \includegraphics[width=\textwidth]{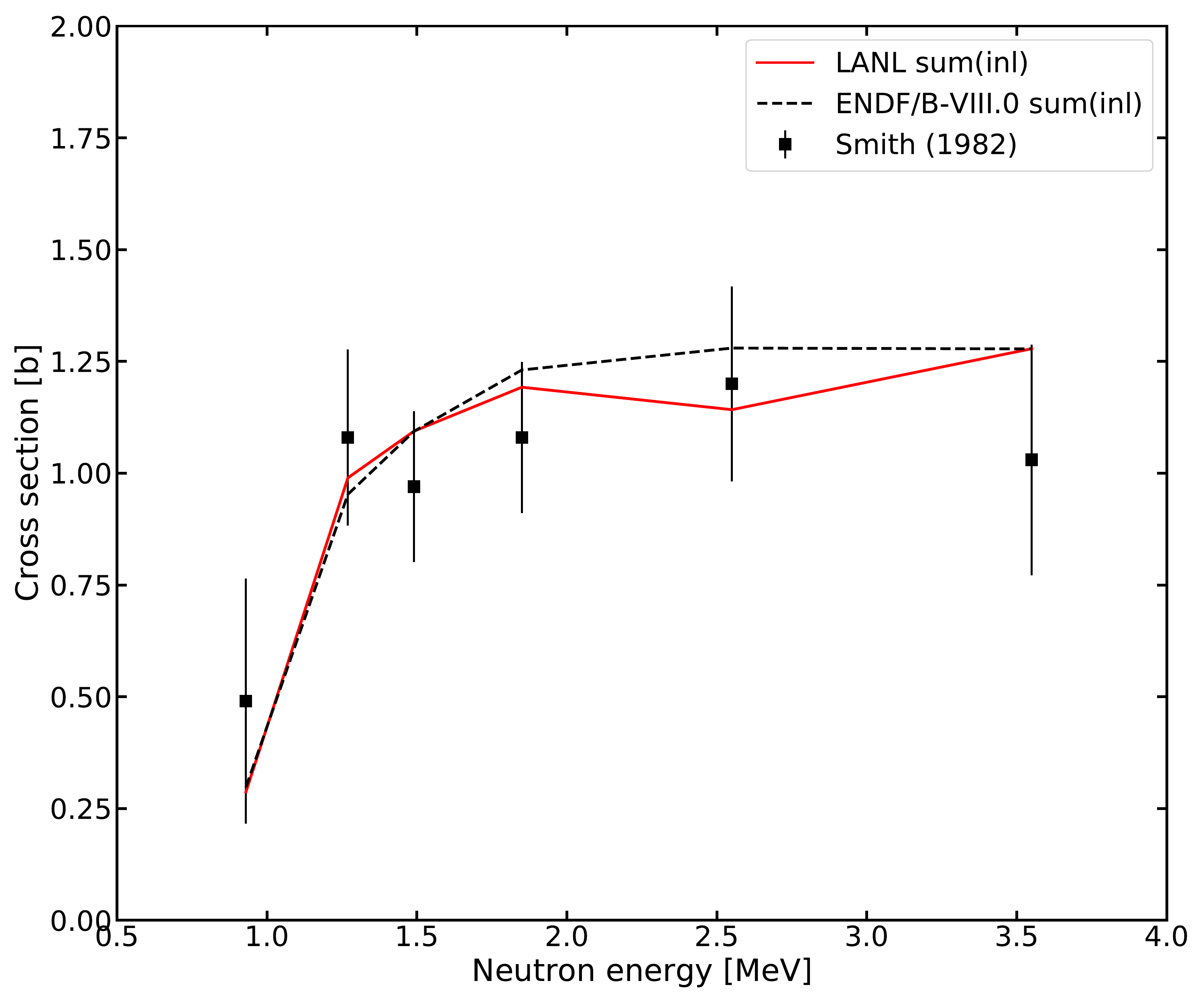}
 \caption{The inelastic cross section constrained by level energy, $E_\textrm{x} \leq 200$ keV. }
 \label{fig:cs_inel_angdist1}
 \end{center}
\end{figure}

\begin{figure}
 \begin{center}
 \includegraphics[width=\textwidth]{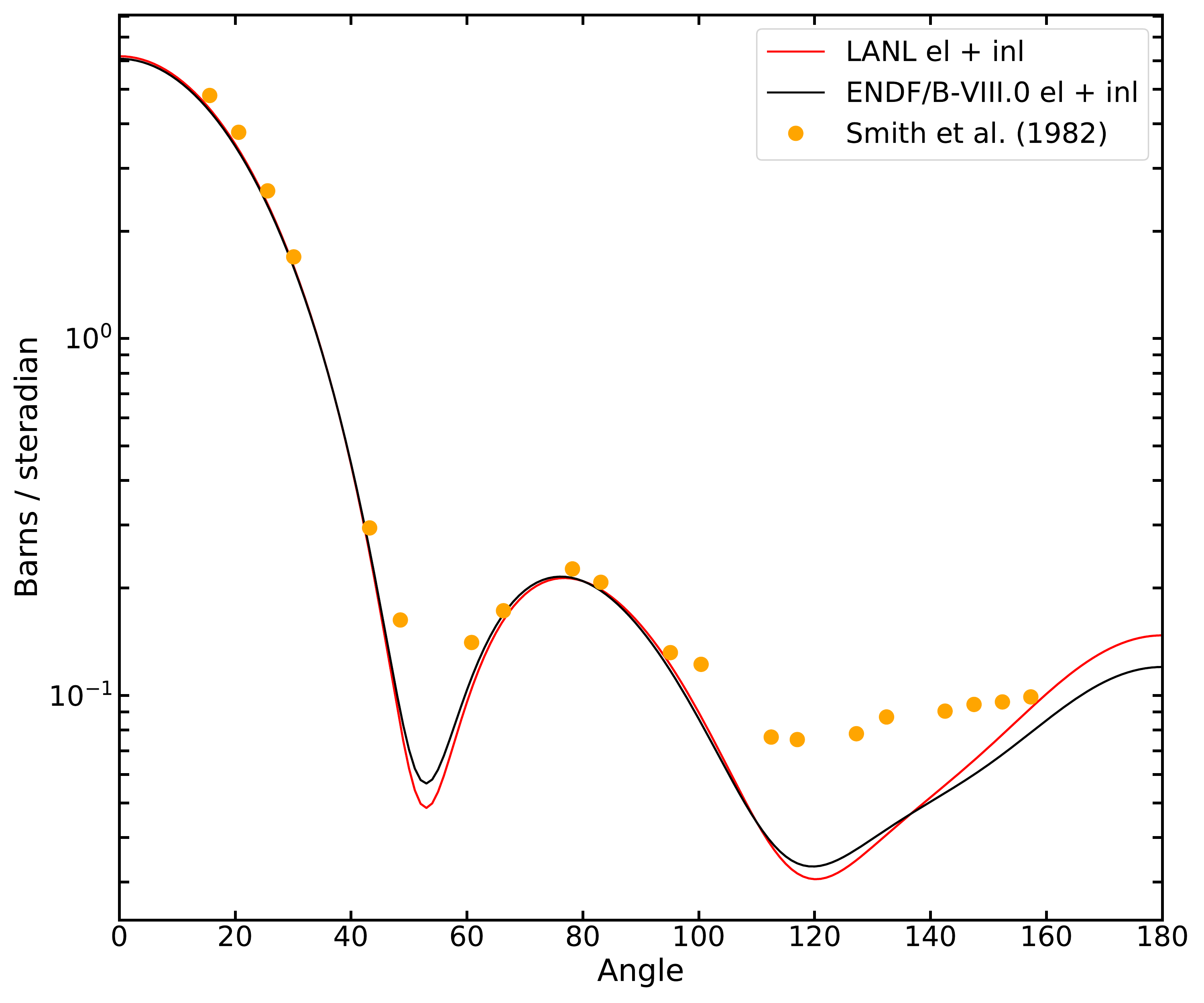}
 \caption{The angular distribution with the inelastic cross section constrained by level energy, $E \leq 200$ keV, as in the previous figure. }
 \label{fig:cs_inel_angdist2}
 \end{center}
\end{figure}

The inelastic cross section constrained by energy level, $E_{x} \leq 200$ keV, is shown in Figure \ref{fig:cs_inel_angdist1}. 
This well reproduces the data of Smith \textit{et al.} (1982) \cite{Smith1982}. 
The angular distribution under the same constraint is shown in Figure \ref{fig:cs_inel_angdist2}. 

\begin{figure}
 \begin{center}
 \includegraphics[width=\textwidth]{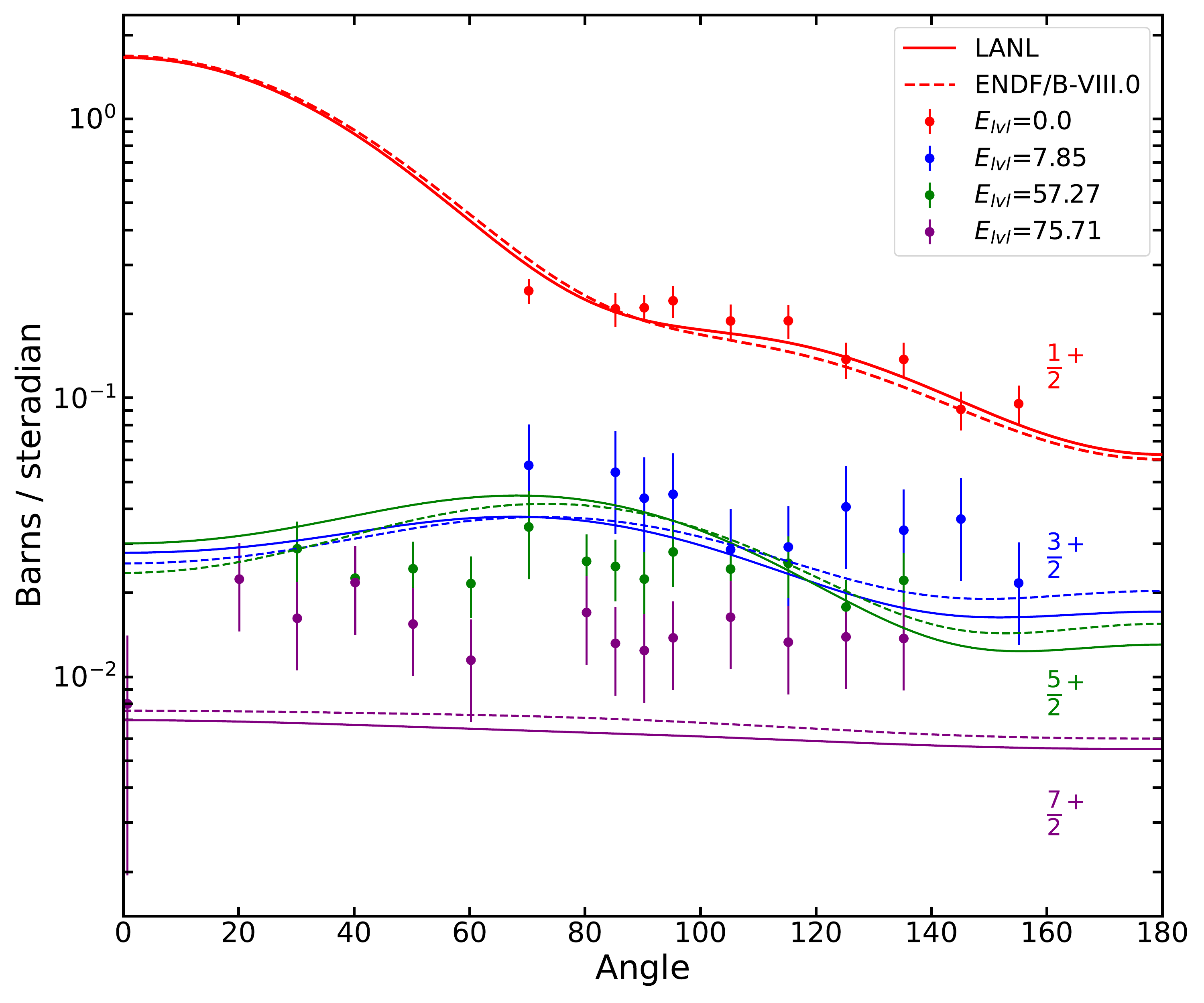}
 \caption{The angular distribution of particular levels in \Pu{} (solid lines - our evaluation) as compared with \ENDF{} (dashed lines). }
 \label{fig:cs_inel_angdist3}
 \end{center}
\end{figure}

The angular distribution of particular levels in \Pu{} are shown in Figure \ref{fig:cs_inel_angdist3}. 
Here the evaluation (solid lines) performs similarly to \ENDF{} (dashed lines). 

\begin{figure}
 \begin{center}
 \includegraphics[width=\textwidth]{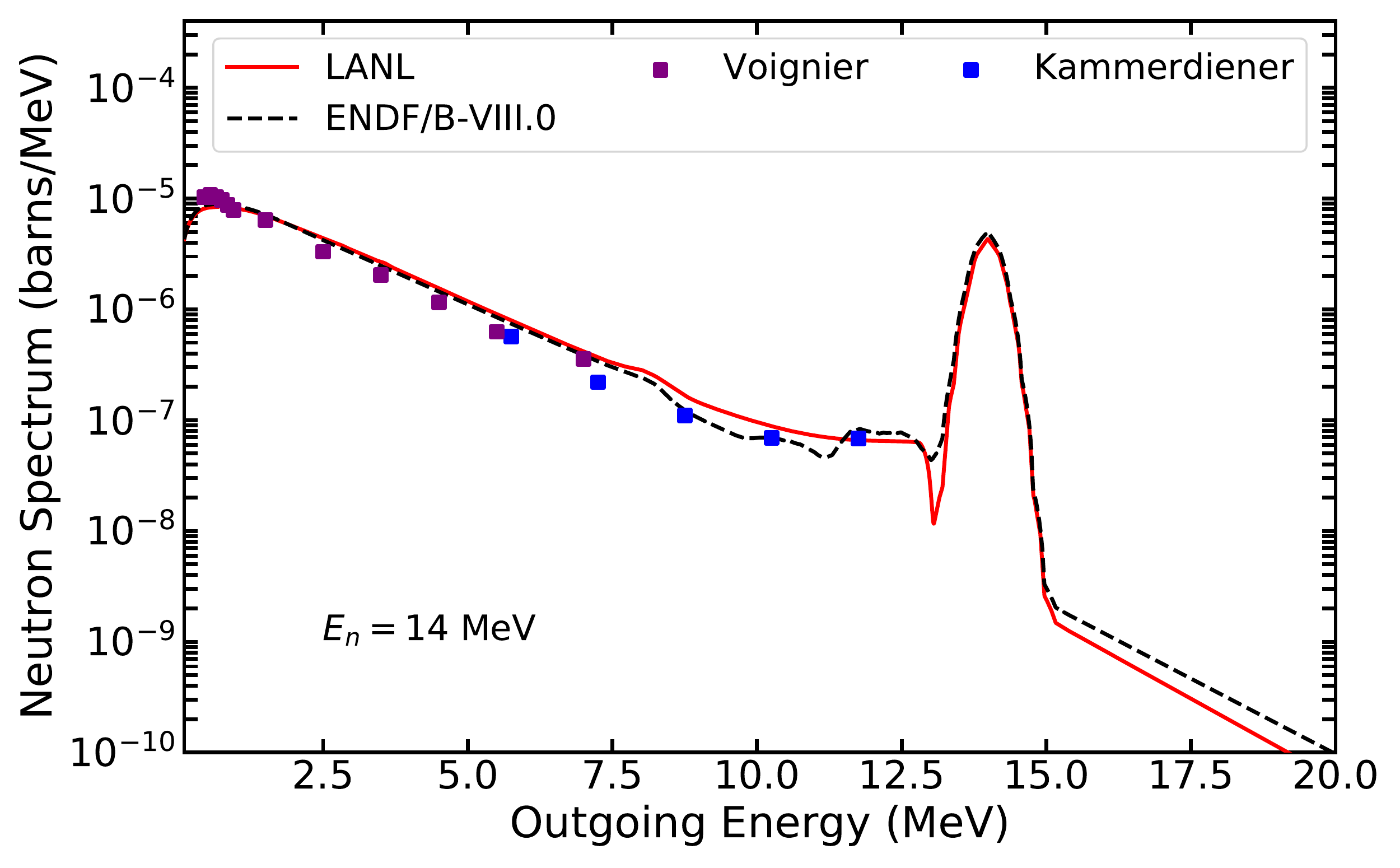}
 \caption{The outgoing neutron spectrum from a fixed incident energy of 14 MeV. }
 \label{fig:nspec_14MeV}
 \end{center}
\end{figure}

The outgoing neutron spectrum of \Pu{}(n,x), where x is any channel is plotted in Figure \ref{fig:nspec_14MeV}. 
Here the evaluation is also on par with \ENDF{} albeit slightly higher than Kammerdiener data between 6 and 11 MeV. 

\section{Radiative Capture}

Since the last evaluation of \Pu{} in \ENDF{}, the most significant experimental development in the capture channel has been the release of the Mosby data \cite{Mosby2014, Mosby2018}. 
In the fast energy range this dataset fills in a dense region of incident energy points extending up to 1.33 MeV~\cite{Mosby2018}. 
The data is in good agreement with that of Hopkins \textit{et al.} (1962) \cite{Hopkins1962} and Gwin \textit{et al.} (1976) \cite{Gwin1976}. 
However, the Mosby data is in tension with Kononov \textit{et al.} (1975) \cite{Kononov1975} between 25 and 700 keV, and we therefore reduce the relative weight of the Kononov dataset in our evaluation. 
The data included in the evaluation are summarized in Table~\ref{tab:capture}. 

\def\arraystretch{1.5}%
\begin{table}
\centering
\caption{Experimental datasets and previous evaluations used in the evaluation of the radiative capture (n,$\gamma$) cross section of \Pu{} that have at least one measurement in the fast energy range. Entries in the table are sorted by increasing year of publication. }
\label{tab:capture}
\begin{tabular}{|c|c|c|c|} 
\hline
\textbf{First Author} & \textbf{Energy Range (MeV)} & \textbf{Year} & \textbf{Reference}  \\ \hline
J.C.~Hopkins & 0.03-1.0 & 1962 & \cite{Hopkins1962} \\ \hline
J.A.~Farrell & 2.003$10^{-5}$-0.4318 & 1970 & \cite{Farrell1970} \\ \hline
M.G.~Schomberg & 0.00015-0.0275 & 1970 & \cite{Schomberg1970} \\ \hline
R.~Gwin & 7.5$10^{-5}$-0.150 & 1976 & \cite{Gwin1976} \\ \hline
V.N.~Kononov & 0.015-0.075 & 1975 & \cite{Kononov1975} \\ \hline
S.~Mosby & 0.00106-1.333 & 2018 & \cite{Mosby2018} \\ \hline
\end{tabular}
\end{table}

The theoretical model of capture includes the M1 scissors mode enhancement to the capture cross section for \Pu{}. 
This idea comes from the study of Ullmann \textit{et al.} (2014) \cite{Ullmann2014} who deduced the M1 scissors strength from the experimental capture cross section as well as the $\gamma$-ray multiplicity distributions measured with the DANCE (Detector for Advanced Neutron Capture Experiment) spectrometer at LANSCE (Los Alamos Neutron Science Center). 
A similar improvement when adding the M1 scissors mode was also reported by Guttormsen \textit{et al.} (2014) for other nearby actinides \cite{Guttormsen2014}. 

We compute the M1 enhancement by adding a small Lorentzian to the M1 $\gamma$-strength function,

\begin{equation}
  f_{\rm M1}(E_\gamma) = 8.67 \times 10^{-8}
  \sigma_{\rm M1} \Gamma_{\rm M1} \nonumber
  \times
  \frac{E_{\rm M1} \Gamma_{\rm M1}}{(E_\gamma^2 - E_{\rm M1}^2)^2 + E_\gamma^2 \Gamma_{\rm M1}^2}
  \quad \ .
  \label{eq:m1sf}
\end{equation}

In this equation, $E_\gamma$ is the energy of the $\gamma$-ray and the other quantities are parameters of the scissors mode.
For the location of M1 scissors, we assume a mass-dependence proportional to $A^{-1/3}$.
We also assume that the oscillation amplitude is proportional to the deformation parameter, $\beta_2$, in the compound nucleus.
From the previous study on $^{238}$U \cite{Ullmann2014}, we have
\begin{equation}
   E_{\rm M1} = 80 |\beta_2| A^{-1/3} \quad \mathrm{MeV} \ ,
   \label{eq:m1beta}
\end{equation}
which is similar to the theoretical prediction of $66 \delta A^{-1/3}$, where $\delta$ is the Nilsson deformation \cite{Bes1984}. 
More information on this procedure can be found in Ref.~\cite{Mumpower2017}.
The M1 enhancement removes the artificial scaling factor used in previous evaluations of the (n,$\gamma$) cross section.
The result is an improvement of the physical description of capture that is based on measured actinide data. 
We summarize the parameters of the $\gamma$-strengths used in Table \ref{tab:params_gsf}. 

\def\arraystretch{1.5}%
\begin{table}
\centering
\caption{The $\gamma$-strength parameters used in the modeling of the capture cross section of \Pu{}. The parameters are given in the function form of a Lorentzian. }
\label{tab:params_gsf}
\begin{tabular}{|c|c|c|c|} 
\hline
\textbf{Multipole} & \textbf{Energy} & \textbf{Width} & \textbf{Sigma}  \\ \hline
E1 & 11.47 & 3.1 & 310.770 \\ \hline
E1 & 14.02 & 5.5 & 401.410 \\ \hline
M1 &  6.59 & 4.0 &   1.540 \\ \hline
M1 &  2.71 & 1.5 &   1.250 \\ \hline
E2 & 10.14 & 3.2 &   6.790 \\ \hline
M2 &  6.60 & 4.0 &   0.001 \\ \hline
E3 & 10.13 & 3.2 &   0.005 \\ \hline
\end{tabular}
\end{table}

\begin{figure}
 \begin{center}
 \includegraphics[width=\textwidth]{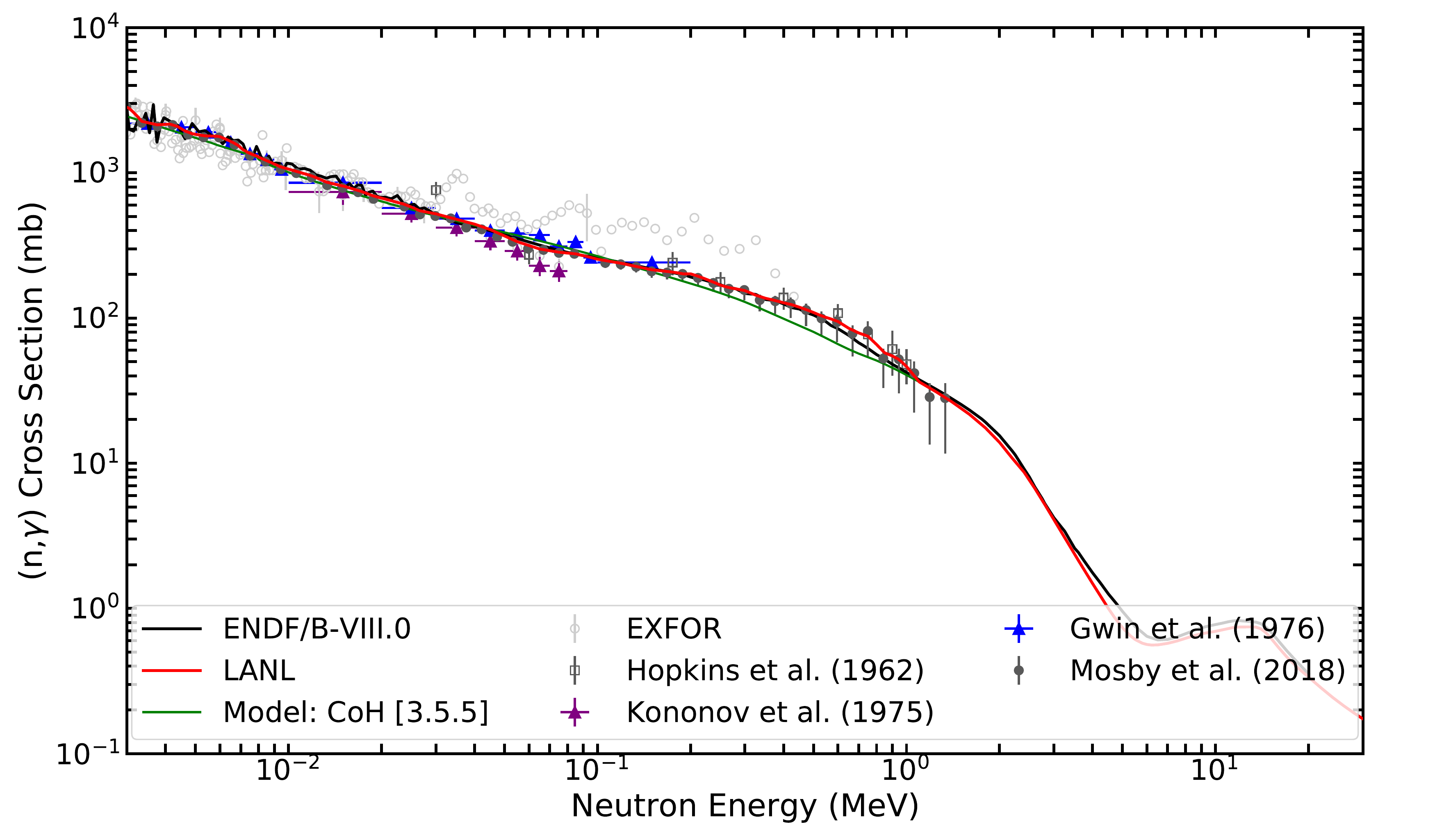}
 \caption{The radiative capture, (n,$\gamma$), cross section above $\sim$ 4 keV. Select datasets are shown in the energy region of the evaluation. }
 \label{fig:cs_ng}
 \end{center}
\end{figure}

The results of our evaluation for capture are shown in Figure~\ref{fig:cs_ng}. 
Capture from the model (green curve) represents a good fit to data, but not of sufficient quality for an evaluation. 
To this end, we use the model as a Bayesian prior and update the cross sections taking into account the high quality data of Mosby, Hopkins and Gwin. 
This produces the final evaluated (n,$\gamma$) cross section (red curve) shown in Figure~\ref{fig:cs_ng}. 

\begin{figure}
 \begin{center}
 \includegraphics[width=\textwidth]{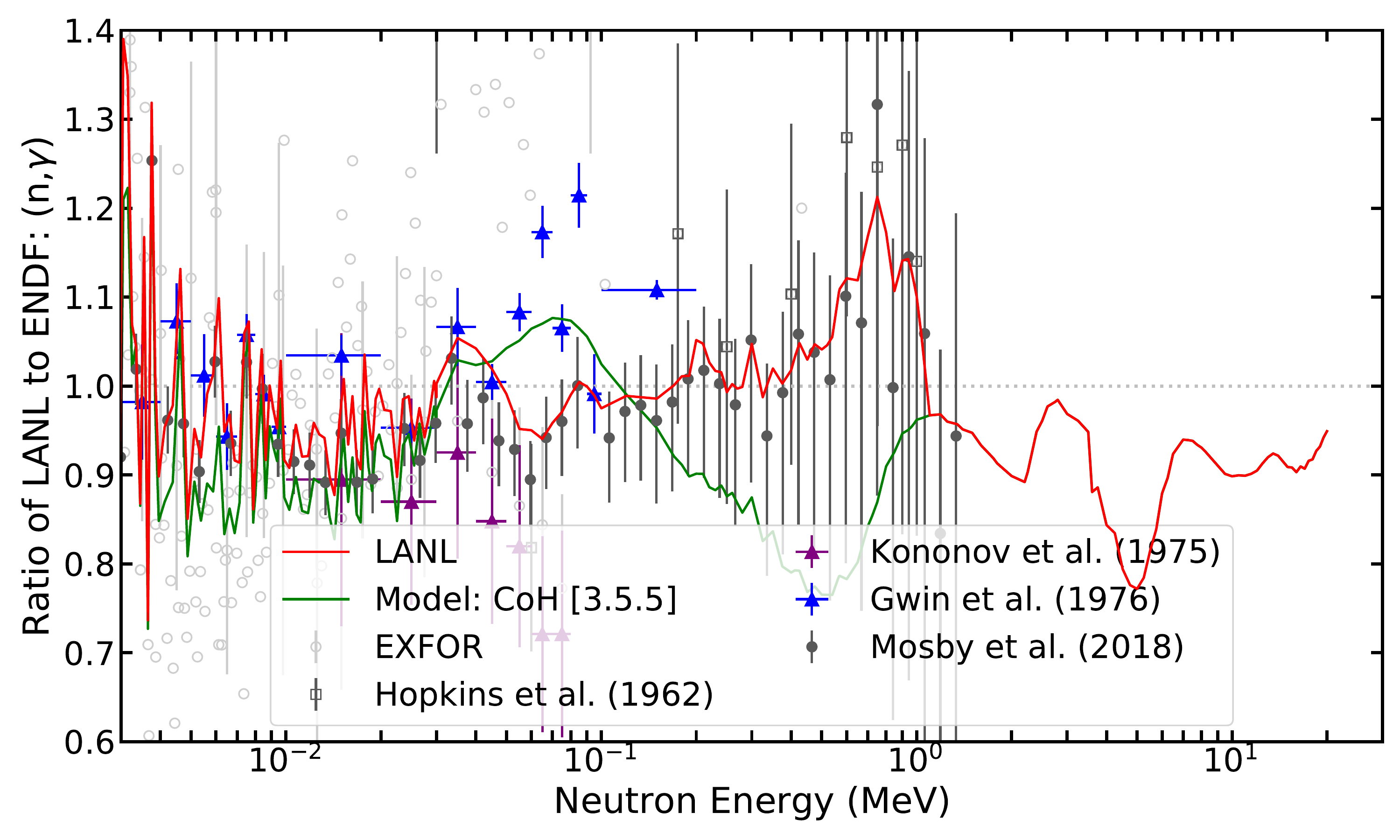}
 \caption{The radiative capture cross section in ratio to \ENDF{}. Our evaluation closely followed the data of Mosby~\cite{Mosby2018}. The jagged nature of our evaluation (red) curve below $\sim$40 keV arises from the artificial oscillations in the \ENDF{} evaluation. }
 \label{fig:cs_ng_ratio}
 \end{center}
\end{figure}

To better understand the changes relative to past work and the \CoH{} model, it is instructive to plot the evaluation in ratio to \ENDF{}. 
As seen in Figure \ref{fig:cs_ng_ratio}, the Mosby data is very influential over nearly the entire energy range. The data of Gwin and Hopkins are influential only in the lower and upper energy ranges respectively. 
The largest changes relative to the \CoH{} model are found in the energy range between 200 keV and 1 MeV. 
This region is approximately where the M1 enhancement has been applied; the model being even more deficient in the region without the enhancement. 
It is therefore of utmost importance to investigate in a similar manner as was done in the case of uranium isotopes (e.g.~see \cite{Ullmann2014}) whether or not an even stronger enhancement may be warranted in this case by looking at associated $\gamma$ multiplicities. 
Future studies in this vein are related to better constraining the static quadruple deformation. 

As can be verified from inspection of Figure \ref{fig:cs_ng}, the evaluated cross section is smooth as a function of incident neutron energy (red curve). 
The same evaluation in Figure \ref{fig:cs_ng_ratio} shows an oscillatory behavior due to the artificial oscillations in the \ENDF{} evaluation below approximately 40 keV. 
Modern data disfavors this oscillatory behavior in preference for a smoother trend, which is exhibited by the current effort. 
Reassuringly, similar behavior is found in other modern evaluations of \Pu{} capture \cite{Capote2022}; this file may be downloaded by following the link in Ref.~\cite{Capote2022zip}. 

\section{(n,2n)}
Despite its importance for applications, relatively large uncertainties remain in the behavior of the (n,2n) cross section across the fast energy range. 
This situation arises due to the high degree of difficulty in performing measurements of this channel. 

The primary datasets considered in our evaluation include the work of Bernstein \textit{et al.} (2002) \cite{Bernstein2002} (GEANIE project), Lougheed \textit{et al.} (2002) \cite{Lougheed2002} and McNabb \textit{et al.} (2001) \cite{McNabb2001}. 
The remainder of the datasets considered in the evaluation of (n,2n) are listed in Table~\ref{tab:n2n}.

Recently, M\'{e}ot and colleagues reported three new data points in the $\sim$7-9 MeV region \cite{Meot2021}. 
The highest incident energy data point was arbitrarily normalized to \ENDF{} at 9.52 MeV which carries an uncertainty of $\pm 15$\%. 
This provides two new data points of interest around 7 MeV. 
Subsequent analysis from Bouland \etal{} questions the behavior of the (n,2n) cross section rise above threshold \cite{Bouland2022}. 
Using statistical model calculations, these authors suggest that the M\'{e}ot \etal{} (2021) \cite{Meot2021} data in fact be re-normalized with a constant factor of 1.24, thus providing an even faster rise from threshold. 
Bouland \etal{} quote a maximum uncertainty on their calculation of $\pm 11.6$ \%. The precise behavior of the cross section in this energy region and at higher incident energies is under continual scrutiny. 

Because data is not sufficiently constraining, the (n,2n) cross section is modeled with the Los Alamos \CoH{} code. 
Statistical model codes tend to overshoot the (n,2n) cross section in the fast energy range. 
We found it was necessary to include a modification to the pre-equilibrium level density in order to evaluate this channel. 
We added a collective enhancement to the 1-particle 1-hole state density in the exciton model. 
The addition of this enhancement results in a decrease to the (n,2n) cross section, while simultaneously allowing for a description of fission \cite{Mumpower2022}. 
The value of the collective enhancement parameter was set to 10 in accordance with pulsed spheres integral data. 
The results of our pulsed spheres simulations are discussed in Section~\ref{sec:benchmarks}. 
No other model parameter adjustments were required to match available data. 

\def\arraystretch{1.5}%
\begin{table}
\centering
\caption{Experimental datasets and previous evaluations used in the evaluation of the (n,2n) cross section of \Pu{} that have at least one measurement in the fast energy range. Entries in the table are sorted by increasing year of publication. }
\label{tab:n2n}
\begin{tabular}{|c|c|c|c|} 
\hline
\textbf{First Author} & \textbf{Energy Range (MeV)} & \textbf{Year} & \textbf{Reference}  \\ \hline
D.S.~Mather    & 6.5-9.0    & 1972 & \cite{Mather1972} \\ \hline
J.~Frehaut     & 6.49-13.09 & 1986 & \cite{Frehaut1986} \\ \hline
L.A.~Bernstein & 6.5-21.96  & 2002 & \cite{Beacker2001, Bernstein2002} \\ \hline
McNabb         & 13.8-14.8  & 2001 & \cite{McNabb2001} \\ \hline
R.W.~Lougheed  & 13.8-14.8  & 2002 & \cite{Lougheed2002} \\ \hline
V.~M\'{e}ot    & 7.1-9.3    & 2021 & \cite{Meot2021} \\ \hline
\end{tabular}
\end{table}

\begin{figure}
 \begin{center}
 \includegraphics[width=\textwidth]{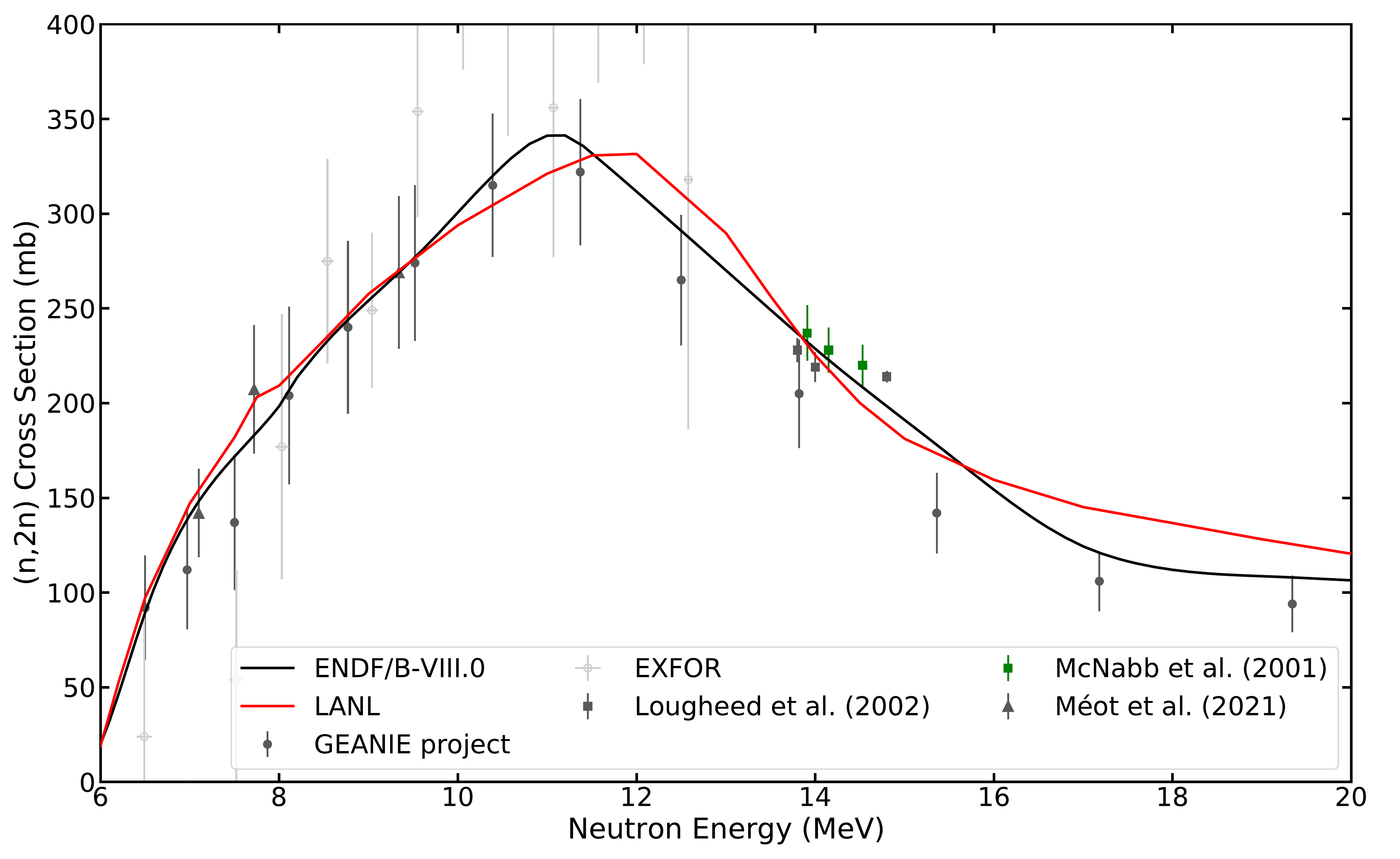}
 \caption{The (n,2n) cross section from threshold to 20 MeV. Select datasets are shown in the energy region of the evaluation. }
 \label{fig:cs_n2n}
 \end{center}
\end{figure}

The result of our (n,2n) evaluation procedure is shown in Figure \ref{fig:cs_n2n}. 
Just above threshold, the evaluation favors the newer M\'{e}ot data, and generally tracks \ENDF{}, albeit slightly higher around 8 MeV. 
The peak of (n,2n) shifts in our evaluation relative to \ENDF{}. 
The down turn now occurs near 12 MeV, in conjunction with the rise in second chance fission. 
In contrast, the \ENDF{} evaluation implemented a straight line linear fit from a peak at roughly 11 MeV through the Lougheed and McNabb data. 
While the cross section value remains the same near 14 MeV, the slope in the current evaluation is more negative, providing a stronger decrease as a function of incident energy. 
Taken together, these two changes above 12 MeV exhibit a more physical behavior between competing channels as compared with \ENDF{}. 

\begin{figure}
 \begin{center}
 \includegraphics[width=\textwidth]{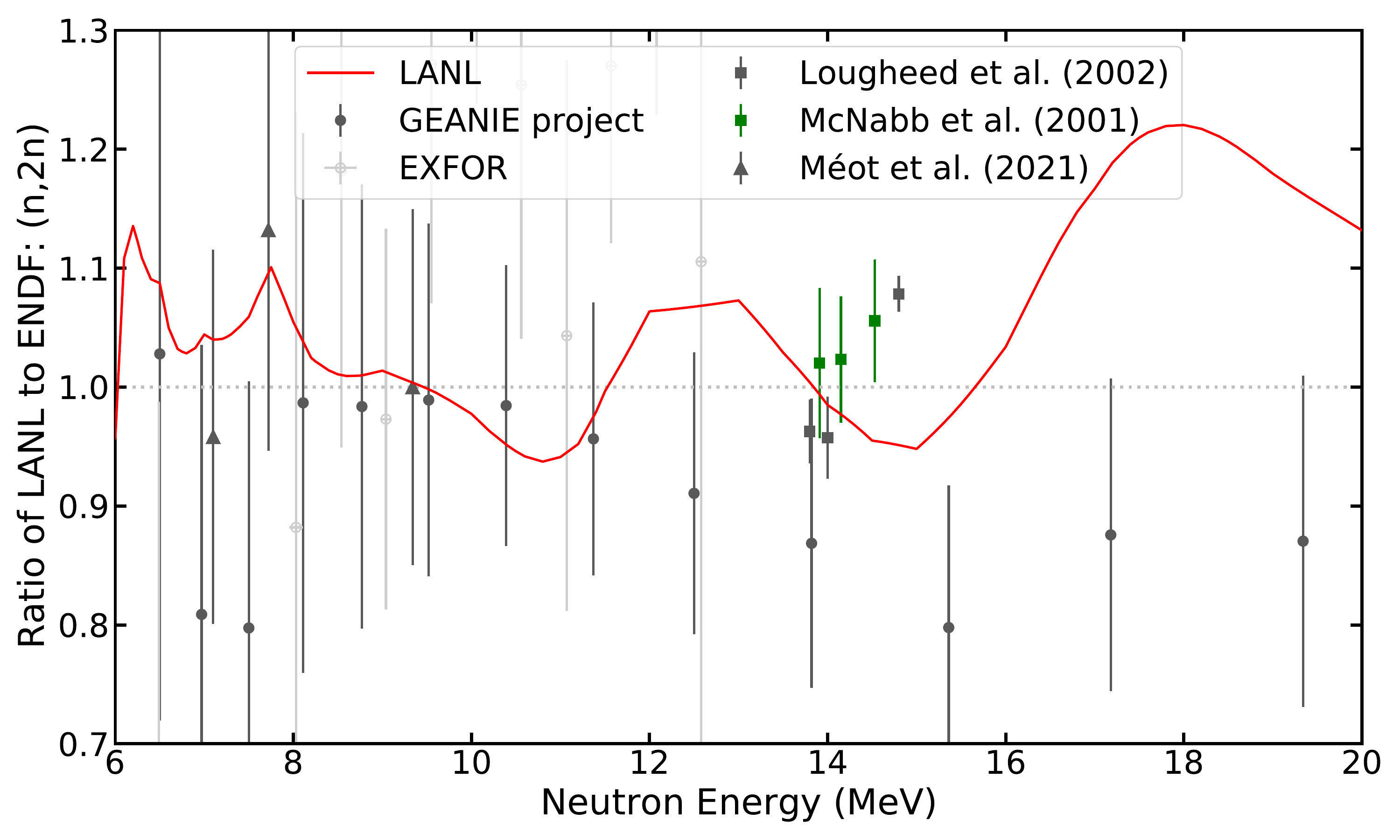}
 \caption{The (n,2n) cross section in ratio to \ENDF{}. Select datasets are shown in the energy region of the evaluation. }
 \label{fig:cs_n2n_ratio}
 \end{center}
\end{figure}

The ratio of the (n,2n) evaluation relative to \ENDF{} is shown in Figure \ref{fig:cs_n2n_ratio}. 
Above threshold to approximately 11 MeV, the evaluation closely tracks GEANIE and M\'{e}ot data. 
At 14 MeV the Lougheed and McNabb data are favored over GEANIE. 
Above 15 Mev, the current evaluation is above GEANIE data. 
However, this dataset can be tracked better with further parameter refinement. 

\section{Prompt Fission Neutron Multiplicity ($\overline\nu_p$)}\label{sec:nubar}

\begin{figure}
 \begin{center}
 \includegraphics[width=\textwidth]{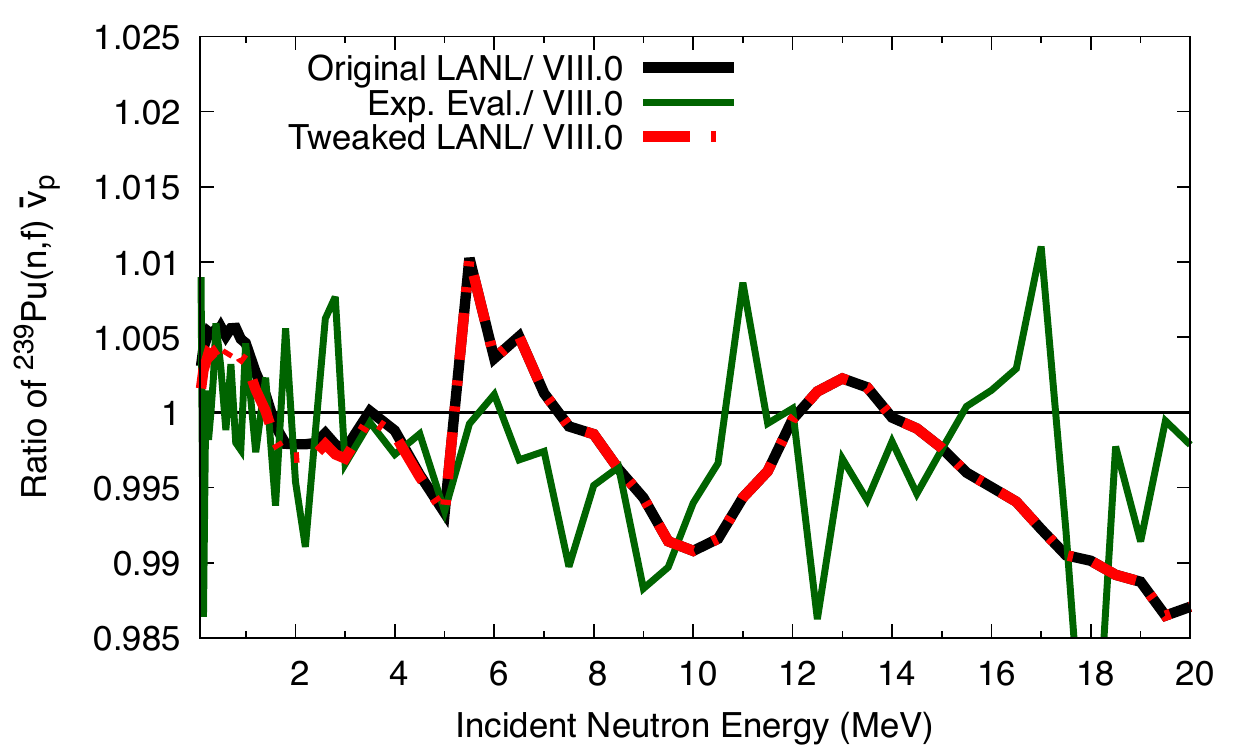}
 \caption{The $\overline\nu_p$ from the current evaluation effort in ratio to \ENDF{}. }
 \label{fig:nubar_ratio}
 \end{center}
\end{figure}

The average prompt fission neutron multiplicity ($\overline\nu_p$) is shown in ratio to \ENDF{} in Figure~\ref{fig:nubar_ratio}. 
The black curve represents the evaluation effort of Neudecker, Lovell and Talou \cite{Neudecker2021nu}; it includes a complete new uncertainty quantification of past experimental data, a new experiment measured by a CEA/ NNSA collaboration~\cite{Marini2021} and modeling via the CGMF fission event generator \cite{Talou2021}.
The green curve shows an evaluation based on only experimental data that allows us to study the impact of including the model in the evaluation. 
The red curve represents the modification enacted to improve the prediction of PU-MET-FAST and PU-MET-INT ICSBEP $k_\mathrm{eff}$ values. 

\begin{figure}
 \begin{center}
 \includegraphics[width=\textwidth]{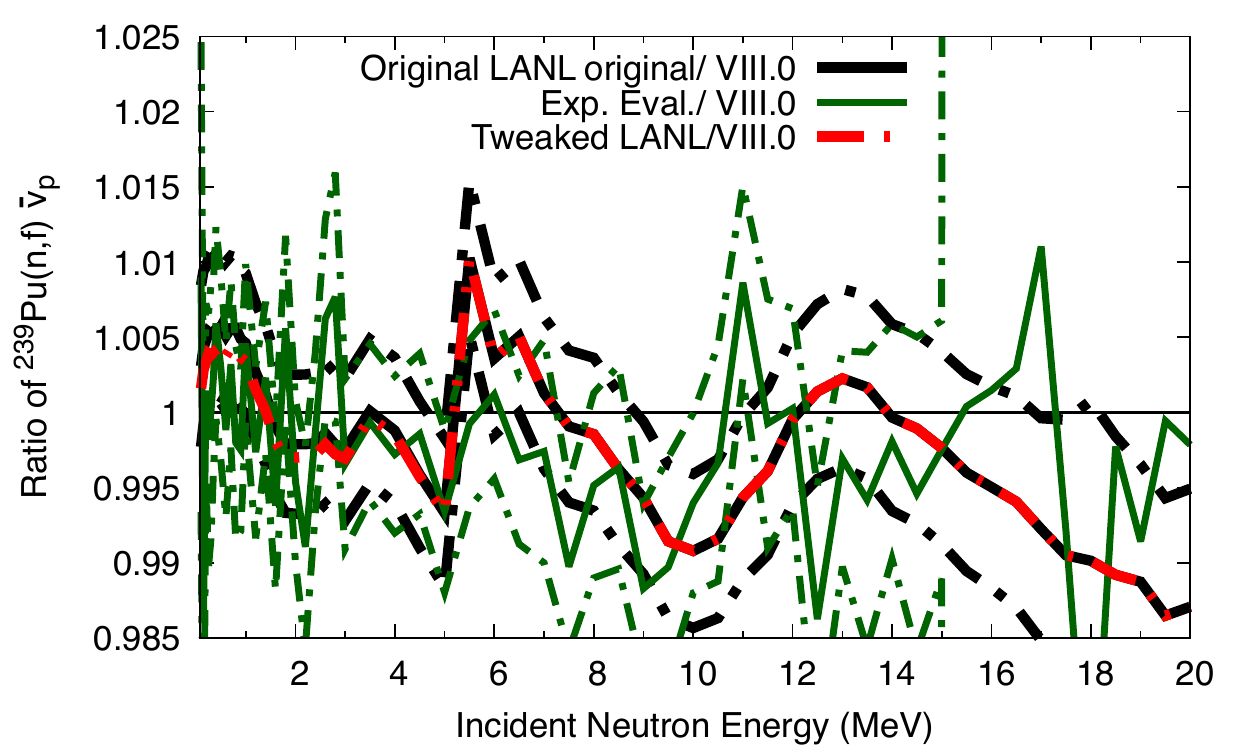}
 \caption{The $\overline\nu_p$ from the current evaluation effort in ratio to \ENDF{} along with uncertainties. }
 \label{fig:nubar_ratio_err}
 \end{center}
\end{figure}

\begin{figure}
 \begin{center}
 \includegraphics[width=\textwidth]{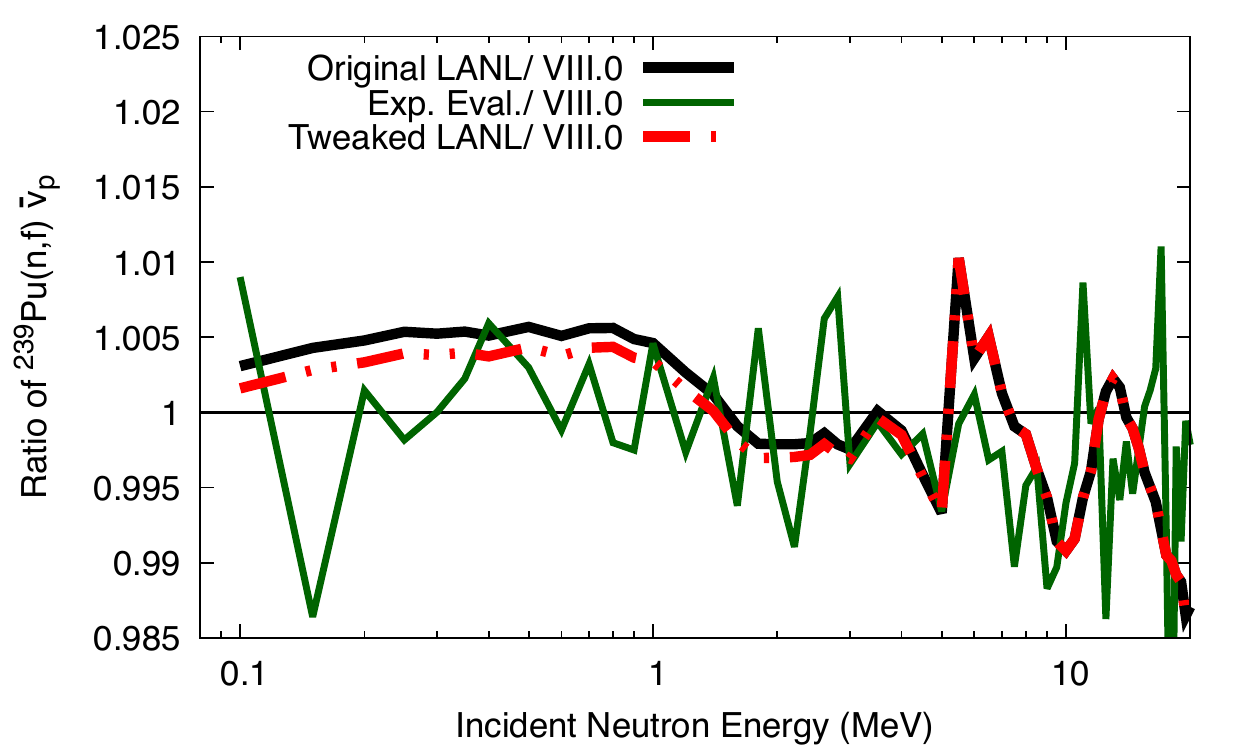}
 \caption{The  $\overline\nu_p$ from the current evaluation effort in ratio to \ENDF{}. The X-axis scale is logarithmic. }
 \label{fig:nubar_ratio_log}
 \end{center}
\end{figure}

\begin{figure}
 \begin{center}
 \includegraphics[width=\textwidth]{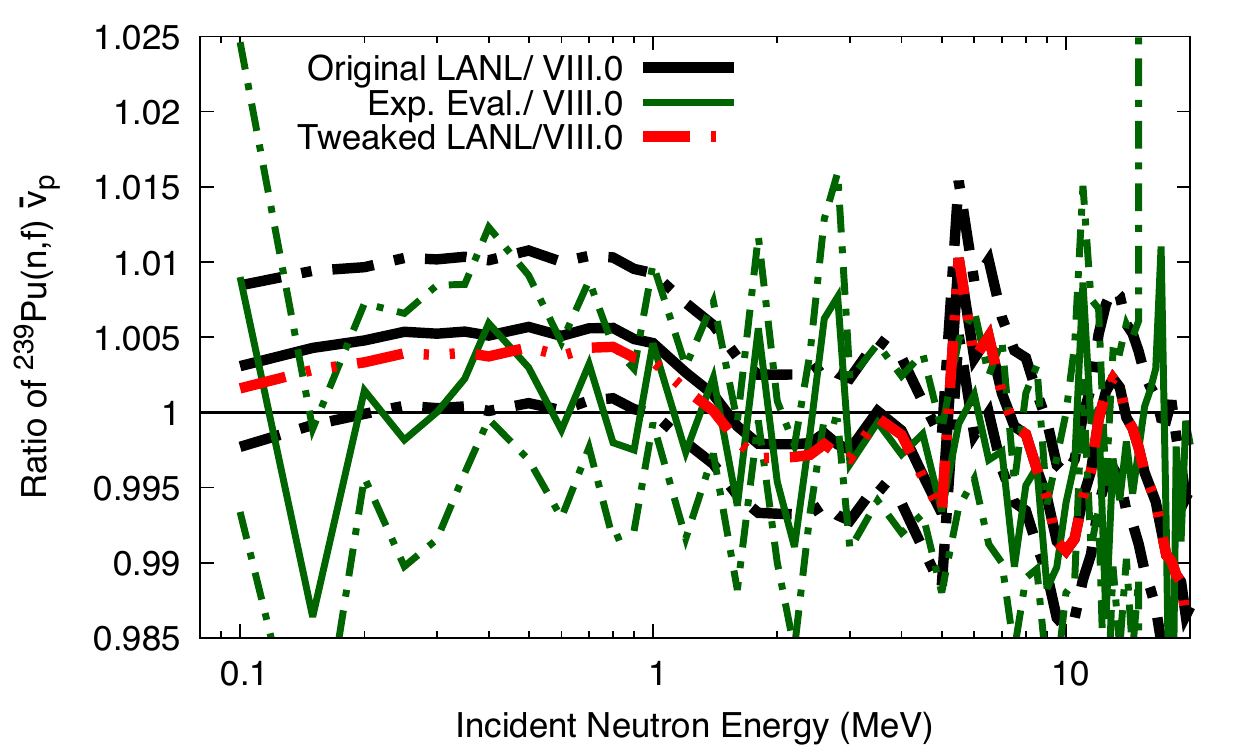}
 \caption{The  $\overline\nu_p$ from the current evaluation effort in ratio to \ENDF{} along with uncertainties. The X-axis scale is logarithmic. }
 \label{fig:nubar_ratio_log_err}
 \end{center}
\end{figure}

This modification linearly decreases the $\overline\nu_p$ as a function of energy, with the largest change occurring at the end of the resonance range, where no constraining experimental data were found and is zero at 5 MeV. 
This change is well within $\overline\nu_p$ uncertainties, as can be verified from Fig.~\ref{fig:nubar_ratio_err}. 
A log-scale of these figures are repeated in Figs.~\ref{fig:nubar_ratio_log} and \ref{fig:nubar_ratio_log_err} respectively. 

\section{Integral Benchmarks}\label{sec:benchmarks}
Validation results for two different evaluations are discussed and were taken from Ref.~\cite{Neudecker2022Val}:
\begin{enumerate}
\item ENDF/B-VIIII.0~\cite{Brown2018}, and
\item ENDF/B-VIII.0 except for the new $^{239}$Pu LANL file from 4/21/2022.
\end{enumerate}
This new file contains the cross section and angular distributions described above.
The PFNS was taken from Refs.~\cite{Neudecker2022PFNS} for 0.5--30 MeV; at thermal, the INDEN evaluation was used.
The (n,f) cross section includes updates from Refs.~\cite{Neudecker2020, Neudecker2021} (including experimental covariances updated according to templates of expected measurement uncertainties and fissionTPC data).
A modified $\overline\nu_p$ from Ref.~\cite{Neudecker2021nu} was adopted where there is little constraining experimental data.
Otherwise, ENDF/B-VIII.0 $^{239}$Pu was carried over unchanged from ENDF/B-VIII.0.
All validation results were obtained by running MCNP-6.2~\cite{MCNP6.2}.
ENDF/B-VIII.0 was used for all other isotopes than $^{239}$Pu.

It should be mentioned that ENDF/B-VIII.0 ACE files of 2020 were used including those for thermal-scattering-law kernels.
This choice was taken to keep consistency for validation results that started to be calculated in 2020.
However, some ACE files changed since then, especially, those for thermal-scattering-law kernels.
Default ACE files will be used for ENDF/B-VIII.0 once this file is delivered to CSEWG.

\subsection{Neutron Multiplication Factor of Selected Fast ICSBEP Critical Assemblies}

\begin{figure}
\centering
\includegraphics[width=0.91\textwidth]{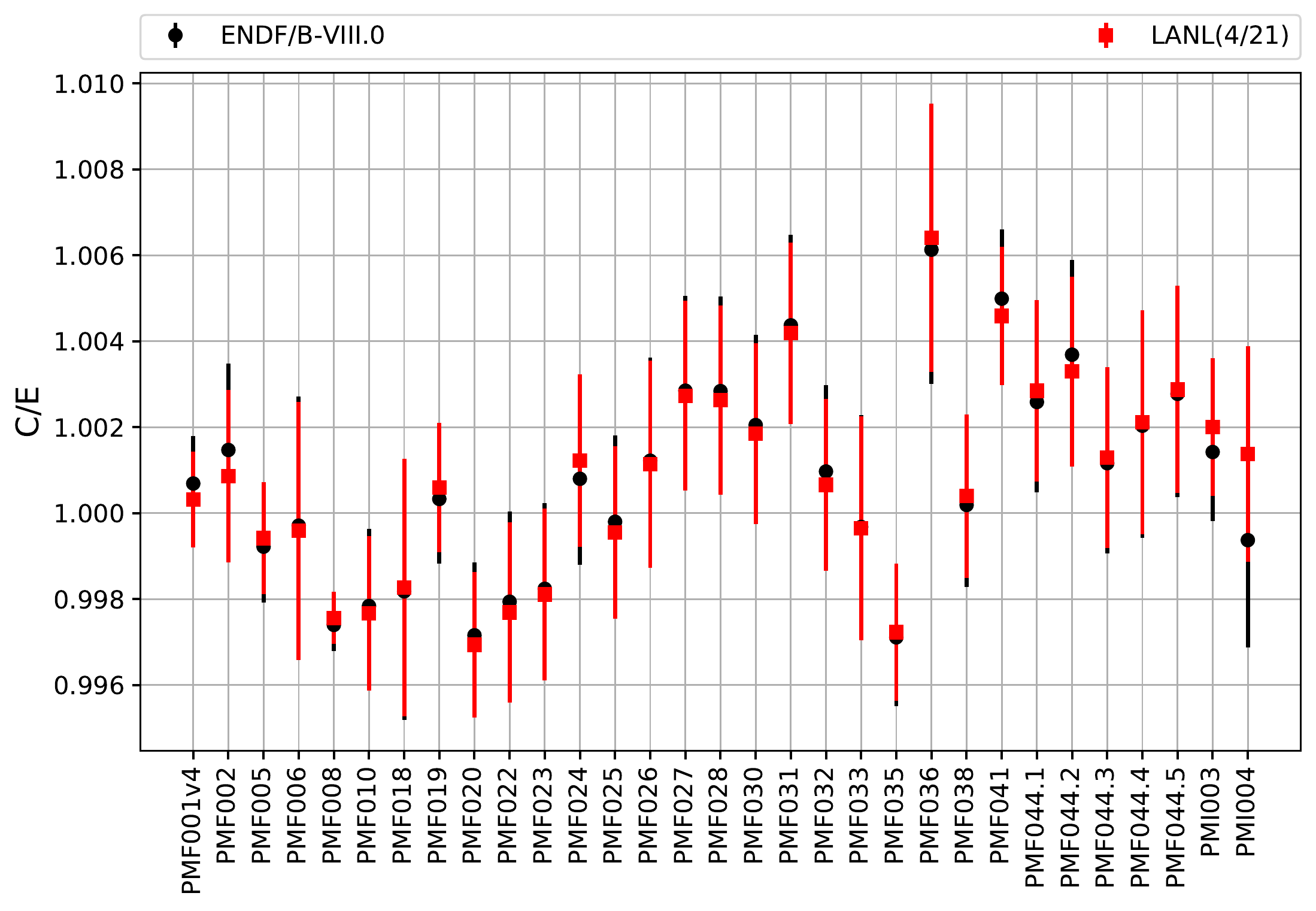}
\includegraphics[width=0.91\textwidth]{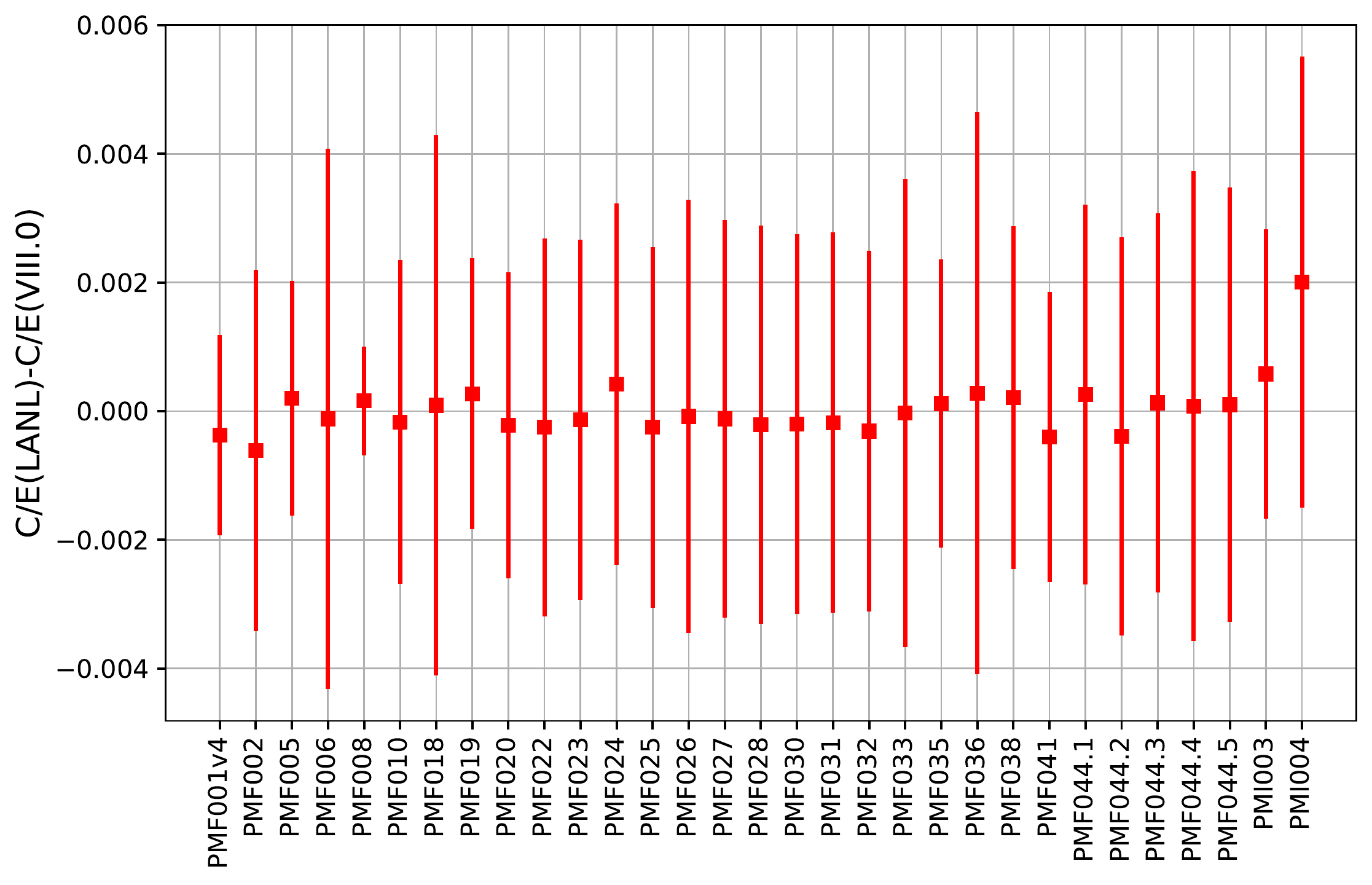}
\caption{Simulated and experimental $k_\mathrm{eff}$ values for ICSBEP critical assemblies are compared using ENDF/B-VIII.0, and ENDF/B-VIII.0 except for the new $^{239}$Pu LANL file.}
\label{fig:keff}
\end{figure}

It is shown in Fig.~\ref{fig:keff} that simulated values of the effective neutron multiplication factor, $k_\mathrm{eff}$, of ICSBEP critical assemblies~\cite{ICSBEP} using the new $^{239}$Pu file are close to ENDF/B-VIII.0 values for most benchmarks; this is reflected also in a similar mean bias across $k_\mathrm{eff}$ values simulated with these two evaluated files: The mean bias is 78 pcm for ENDF/B-VIIII.0, while it is 81 pcm for the LANL file. The mean uncertainty on the bias is approximately 10 pcm. 

The mean bias is slightly higher (but within the Monte Carlo (MC) uncertainties) for the new file, partially owing to the worse prediction of PMI003 and PMI004 $k_\mathrm{eff}$ values.
However, $k_\mathrm{eff}$ values of critical assemblies with comparably harder spectra (PMF001v4, PMF002, PMF005, PMF006, etc.) are better described with the new file than ENDF/B-VIII.0.
Notably, the difference between simulated Jezebel (PMF001v4) and Flattop $k_\mathrm{eff}$ values got smaller by 25 pcm for the new file compared to \ENDF{}.

\subsection{Effective Delayed Neutron Fraction of Selected Fast Plutonium ICSBEP Critical Assemblies}

The effective neutron multiplication factor is one integral quantity that is frequently used to validate nuclear data. Another integral quantity that can be examined is the effective delayed neutron fraction, $\beta_{\textrm{eff}}$. The effective delayed neutron fraction has significance in criticality experiments and nuclear reactor applications. This integral quantity describes the fraction of delayed neutrons that contribute to the criticality of the system. This quantity can be calculated with
\begin{equation}
    \beta_{\textrm{eff}} = 1-\frac{k_p}{k}\textrm{,}
    \label{Eq:basic_beff_eq}
\end{equation}
where $k$ is the neutron multiplication factor, and $k_p$ is the prompt neutron multiplication factor. This quantity was calculated with MCNP6.2~\cite{MCNP6.2} using the KOPTS card. The calculated nuclear data library results presented in Table \ref{tab:beta_eff} are produced using this method. The measured values are retrieved from the International Reactor Physics Evaluation Project (IRPhEP). The results are presented for two criticality experiments: (1) Jezebel, and (2) Flattop with plutonium core. These two benchmarks have the International Criticality Safety Benchmark Evaluation Project (ICSBEP) handbook denotation of PU-MET-FAST-001v4 and PU-MET-FAST-006, respectively. The results in Table \ref{tab:beta_eff} show that the LANL $^{239}$Pu evaluation is in better agreement with measured values than ENDF/B VIII.0. 

\begin{table}[]
    \centering
    \begin{tabular}{|c|c|c|c|}
        \hline
         &  & Calculated & Calculated \\
        Benchmark & Measured & ENDF/B VIII.0 & LANL \\ \hline
        PU-MET-FAST-001v4 (Jezebel) & 0.00195(19) & 0.00183(0)* & 0.00185(0)* \\ \hline
        PU-MET-FAST-006 (Flattop, Pu Core) & 0.00276(23) & 0.00284(4) & 0.00275(4) \\ \hline
    \end{tabular}
    \caption{Calculated and experimental values of effective delayed neutron fraction for particular fast plutonium benchmark experiments. The denotation of * was used to signify that the uncertainty of the calculated result is less than 1 pcm. Only experimental uncertainties were provided for both assemblies, which were unrealistically small. Hence, 8\% uncertainties were added to experimental ones in quadrature following IRPhEP.}
    \label{tab:beta_eff}
\end{table}

\subsection{Spectral Indexes in Jezebel and Flattop Critical Assemblies}
Spectral indexes in the Jezebel critical assembly were calculated with the two combinations of files, ENDF/B-VIII.0, and ENDF/B-VIII.0 including the LANL file, and are shown in Table~\ref{tab:Validation_keffRR}.
The aim here was to test how close these values are to those predicted with ENDF/B-VIII.0.
Simulated values using the new evaluations are close to ENDF/B-VIII.0 for $^{239}$Pu(n,2n)/$^{239}$Pu(n,f), $^{238}$U(n,f)/$^{235}$U(n,f), $^{233}$U(n,f)/$^{235}$U(n,f), and $^{239}$Pu(n,f)/$^{235}$U(n,f) in Jezebel.

\begin{table*}[htb]
 \caption{Simulated values for $k_\mathrm{eff}$ and spectral indexes of Jezebel (PMF001v4) are compared with each other for ENDF/B-VIII.0, and ENDF/B-VIII.0 with the $^{239}$Pu LANL file. Only MC statistics uncertainties are shown.}
 \label{tab:Validation_keffRR} 
 \centering
 \begin{tabular}{c||c|c}
Observable &	VIII.0 & VIII.0+LANL \\  
\hline
$k_\mathrm{eff} $ & 1.00069(1) & 1.00036(1) \\
$\frac{^{239}\mathrm{Pu} \sigma_{n,2n}}{^{239}\mathrm{Pu}\sigma_{n,f}}$ & 0.00230(5) & 0.00229(8) \\
$\frac{^{239}\mathrm{Pu} \sigma_{n,\gamma}}{^{239}\mathrm{Pu}\sigma_{n,f}}$ & 0.0345(2) & 0.0361(4) \\
$\frac{^{238}\mathrm{U} \sigma_{n,f}}{^{235}\mathrm{U}\sigma_{n,f}}$ & 0.212(1) &  0.211(2) \\
$\frac{^{237}\mathrm{Np} \sigma_{n,f}}{^{235}\mathrm{U}\sigma_{n,f}}$ & 0.9768(5) &  0.9706(8)\\
$\frac{^{233}\mathrm{U} \sigma_{n,f}}{^{235}\mathrm{U}\sigma_{n,f}}$ & 1.566(7) &  1.566(11) \\
x$\frac{^{239}\mathrm{Pu} \sigma_{n,f}}{^{235}\mathrm{U}\sigma_{n,f}}$ & 1.427(6) & 1.424(10)  \\
\end{tabular}
\end{table*}

\subsection{LLNL Pulsed-sphere Neutron-leakage Spectra}

\begin{figure}
\centering
\includegraphics[width=0.49\textwidth]{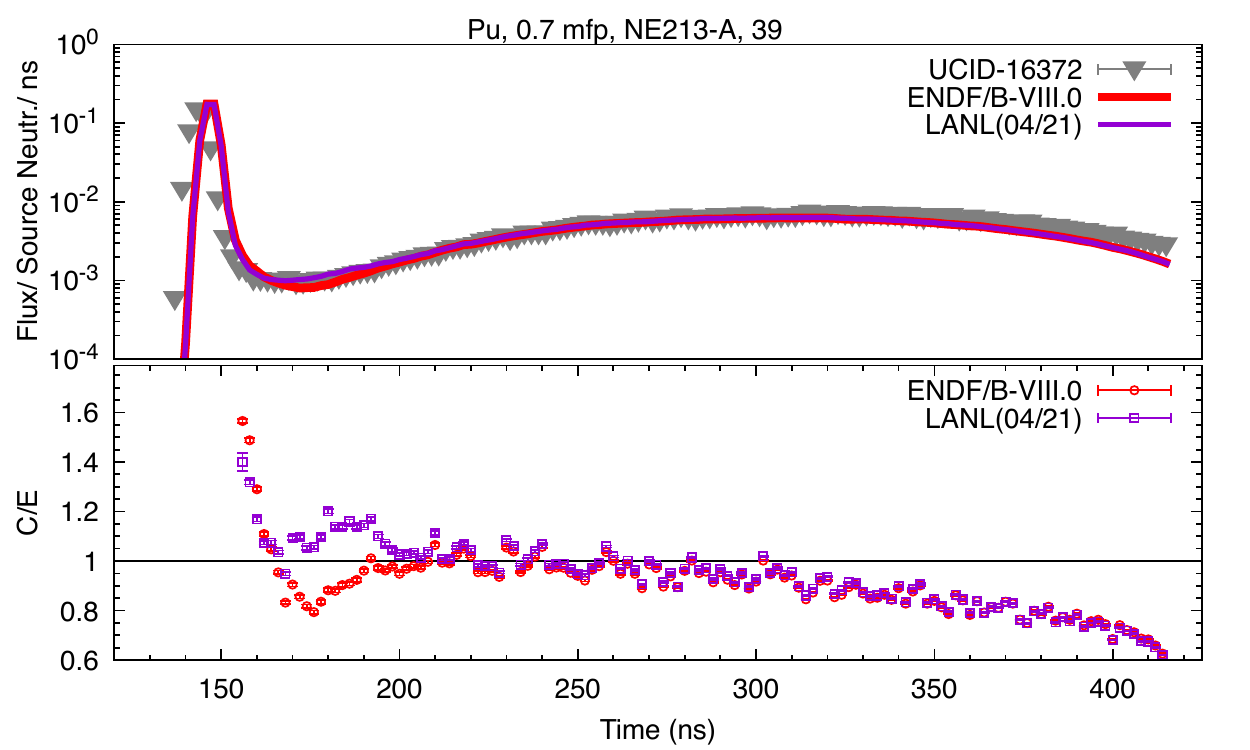} \includegraphics[width=0.49\textwidth]{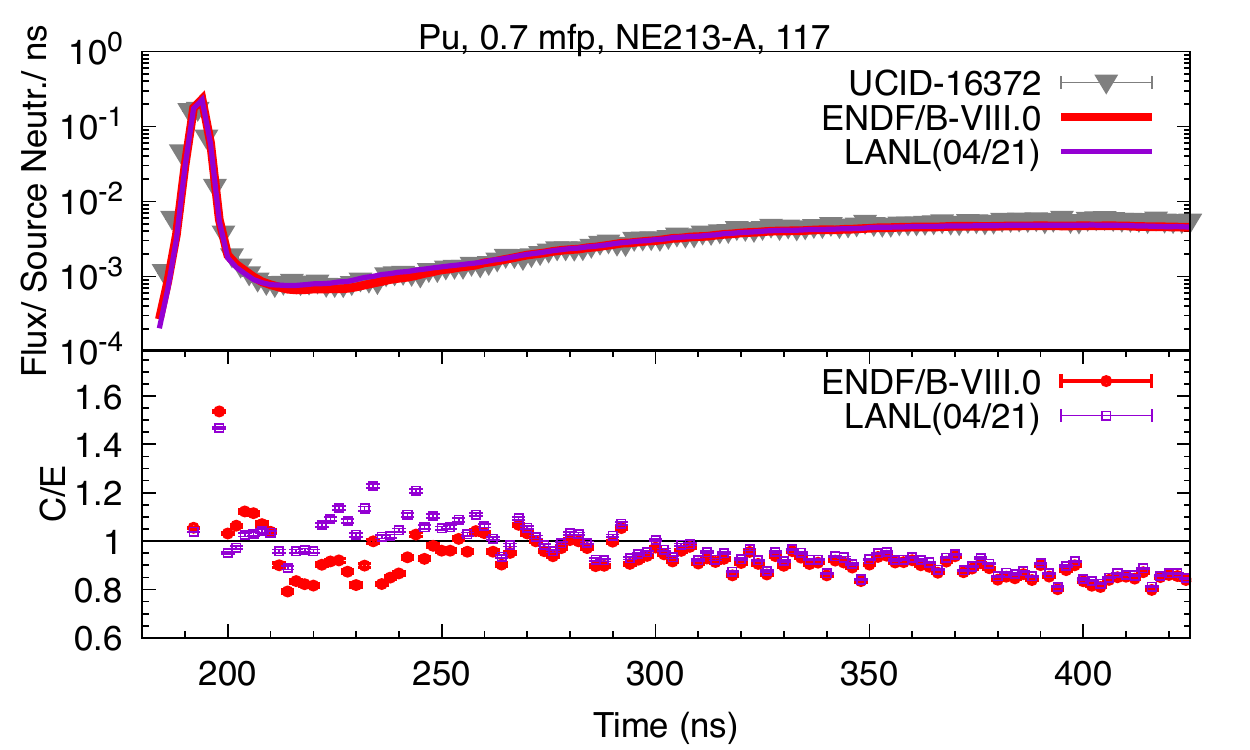} \\
\includegraphics[width=0.49\textwidth]{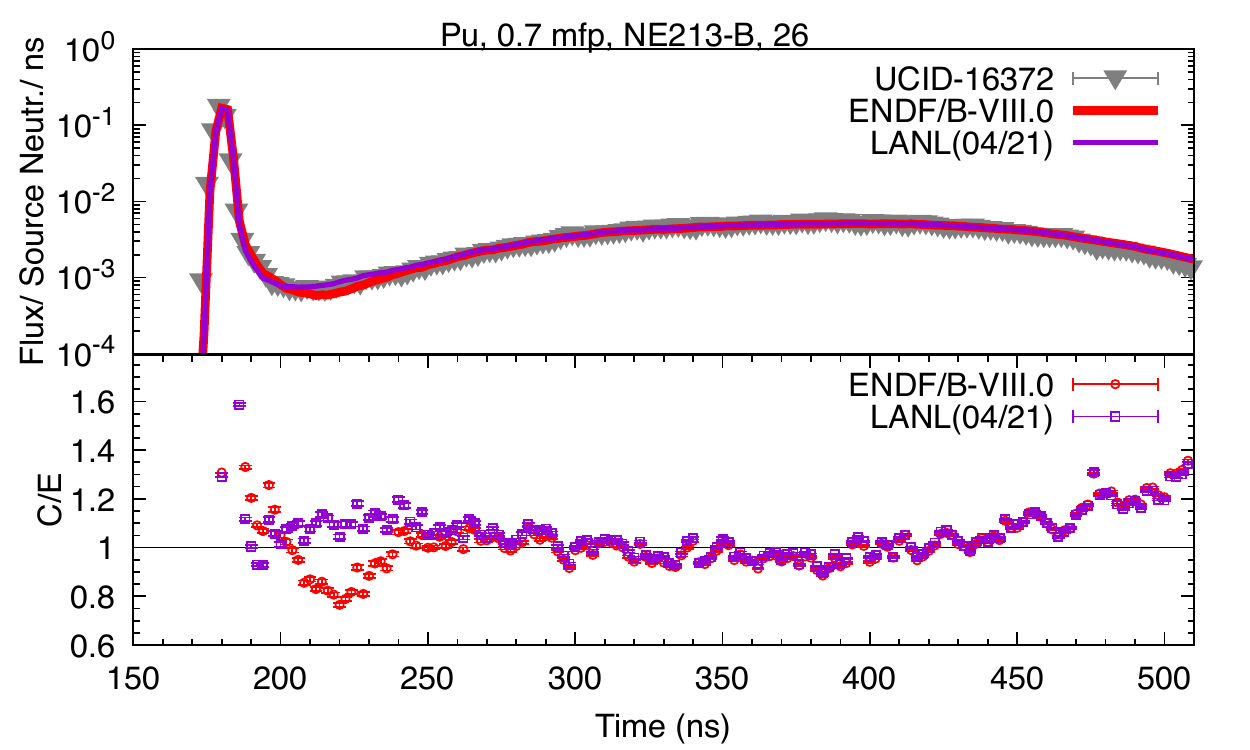}
\caption{Calculated and experimental values are shown for Pu LLNL pulsed spheres. Calculated values are given for ENDF/B-VIII.0 as well as the LANL $^{239}$Pu file.}
\label{fig:Pu9nubarevalValidation_PS}
\end{figure}

LLNL pulsed spheres~\cite{PulseSpheresLLNLsummary} allow us to validate $^{239}$Pu nuclear data from approximately 10--15 MeV.
While the peak of the neutron-leakage spectrum is mostly sensitive to elastic and discrete inelastic scattering, the valley after the peak is strongly influenced by continuum inelastic scattering.
After the valley, fission plays a dominant role, especially the PFNS~\cite{Neudecker:2021_PS}.
The new continuum scattering data lead to an improved description of the valley of the neutron-leakage spectra in Fig.~\ref{fig:Pu9nubarevalValidation_PS}, while the new PFNS lead to negligible changes for times $>$ 270 ns.

\section{Covariances}

The complete covariances may be estimated using the Bayesian fit techniques. 
The computation of covariances in this technique requires not only high performance computing but also sufficient storage space to store all of the parameter changes and model outputs.
These computations are underway. 
In the meantime, an assumption of a linear response to the parameter changes can be made \cite{Kawano2006cov}. 
In this technique, the posterior covariances for parameters take the form,
\begin{equation}
    \label{eqn:cov_posterior}
    \pmb{P} = (\pmb{X}^{-1} + \pmb{C}^{T} \pmb{V}^{-1} \pmb{C}) ^{-1} \ ,
\end{equation}
where $\pmb{P}$ is the prior model parameter covariance matrix, $\pmb{V}$ is the experimental data covariance matrix, and $\pmb{C}$ is the linear response model sensitivity matrix \cite{Rising2013}. 
The final covariance matrix (function of energy) is,
\begin{equation}
    \label{eqn:cov_final}
    \pmb{F} = \pmb{C} \pmb{P} \pmb{C}^{T} \ ,
\end{equation}
where $\pmb{P}$ is the final parameter covariances computed after looping over all datasets. 

For each major reaction channel we show below the variance derived from this procedure, along with the correlation matrix and relative uncertainties. 

\begin{figure}
\centering
\includegraphics[width=0.9\textwidth]{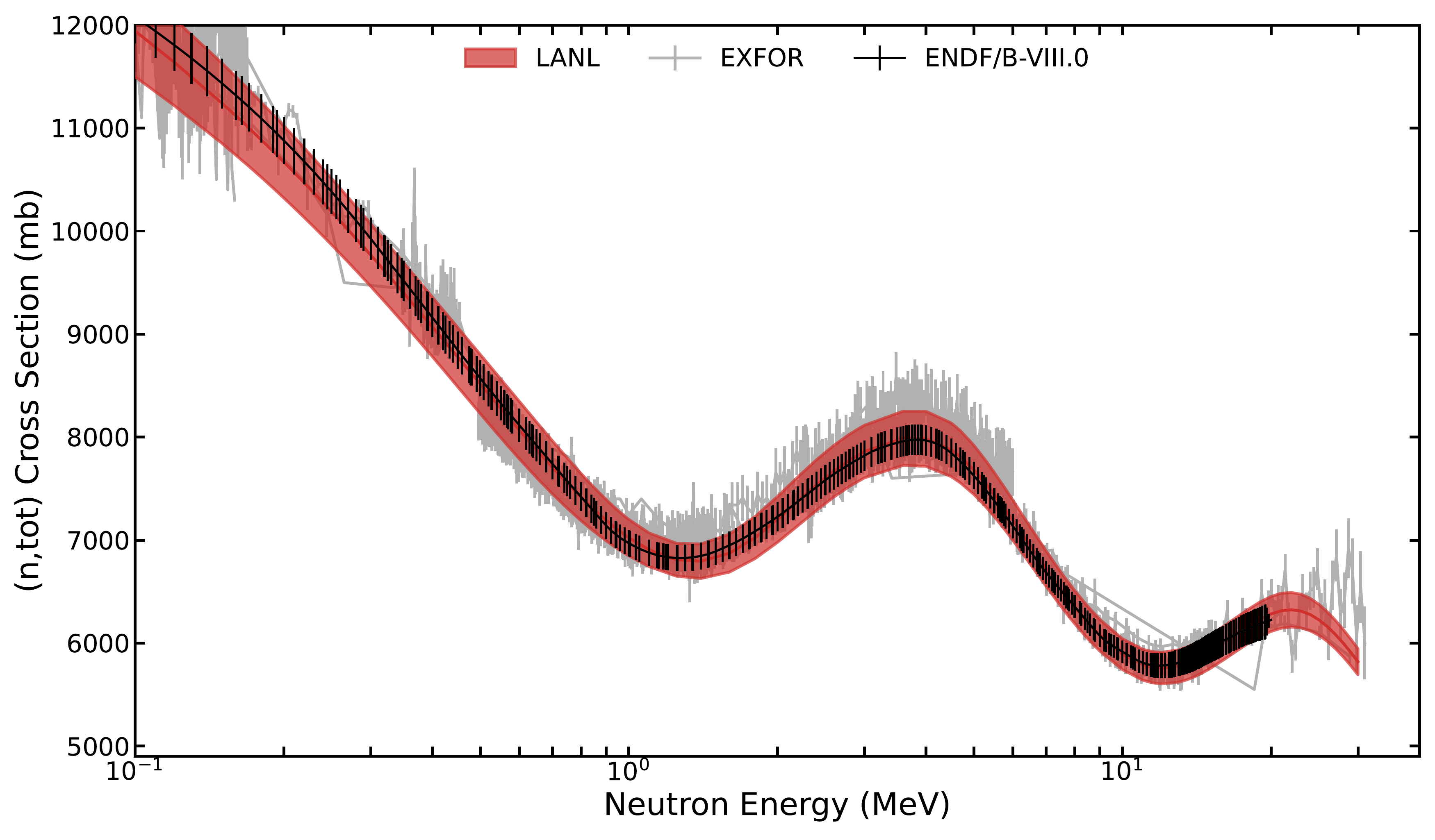}
\caption{Variance for total cross section.}
\label{fig:cov_239pu_cs_tot}
\end{figure}

\begin{figure}
\centering
\includegraphics[width=0.49\textwidth]{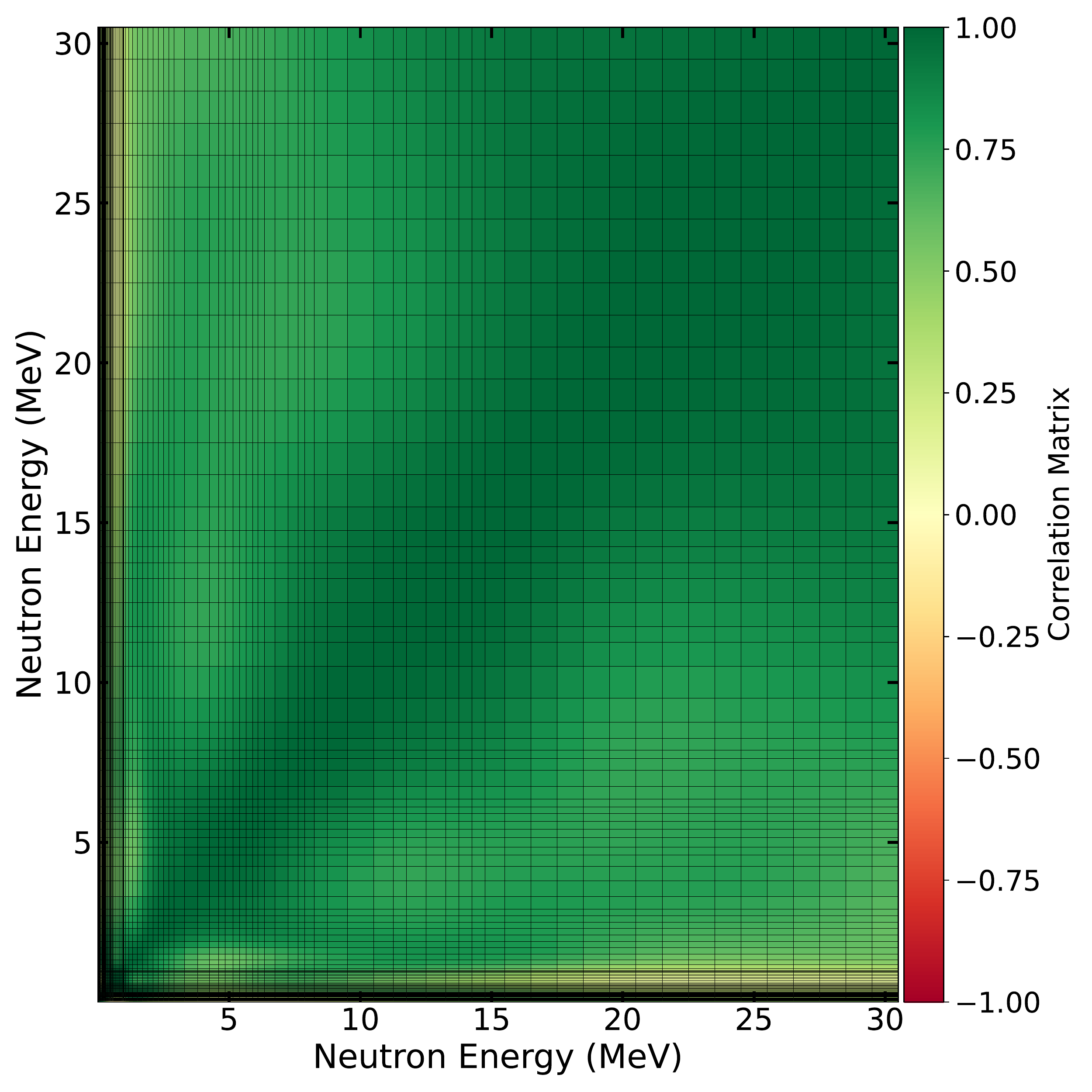}
\includegraphics[width=0.49\textwidth]{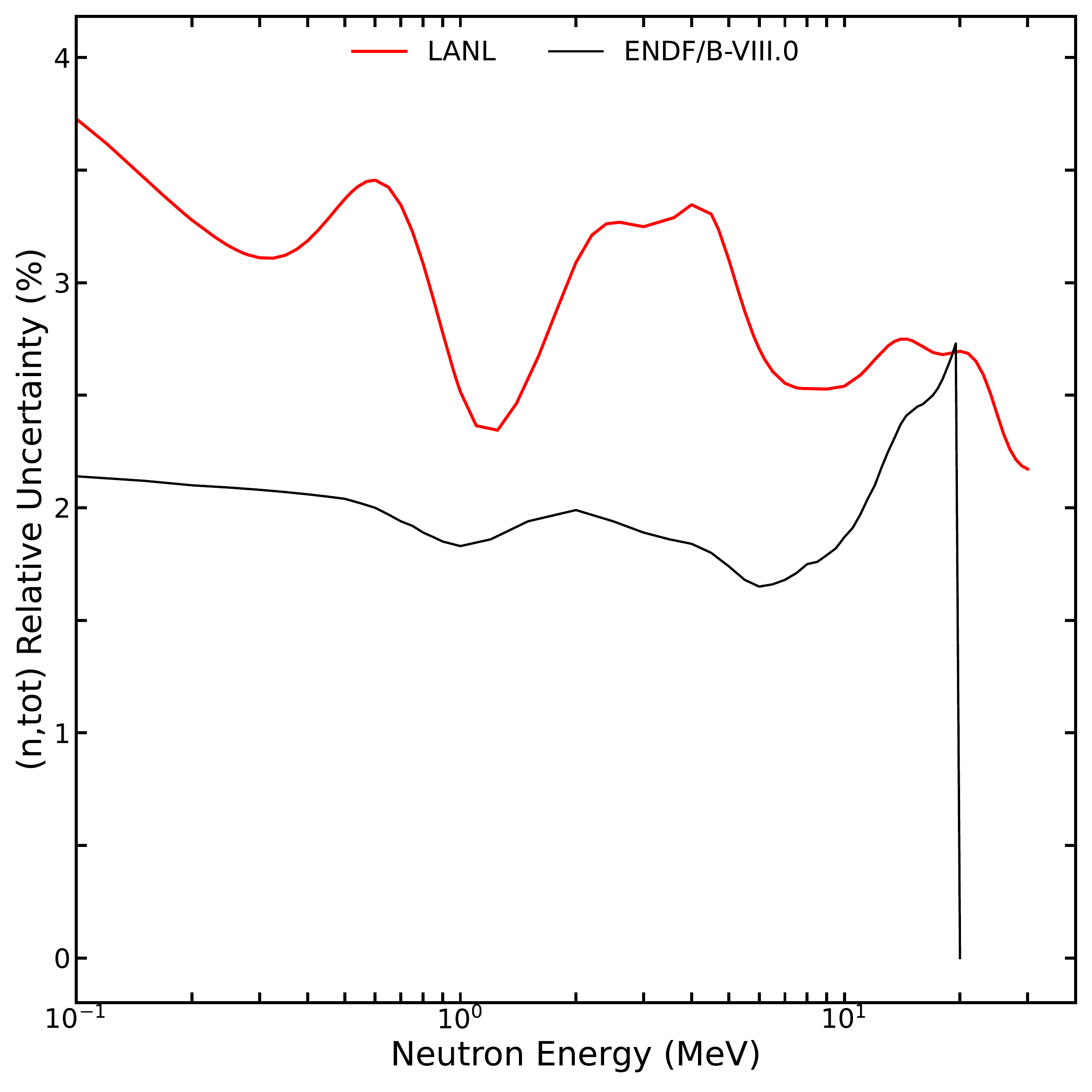}
\caption{(a) Correlation matrix for total cross section. (b) Relative uncertainty as a function of incident energy.}
\label{fig:cov_239pu_tot}
\end{figure}

Figure \ref{fig:cov_239pu_cs_tot} shows the variances (square root of the diagonal entries of the covariance matrix) along the evaluated total cross section. 
Figure \ref{fig:cov_239pu_tot} (a) shows the correlation matrix for the evaluated total cross section and (b) shows the relative uncertainties as a function of incident neutron energy for the evaluated total cross section. 

\begin{figure}
\centering
\includegraphics[width=0.9\textwidth]{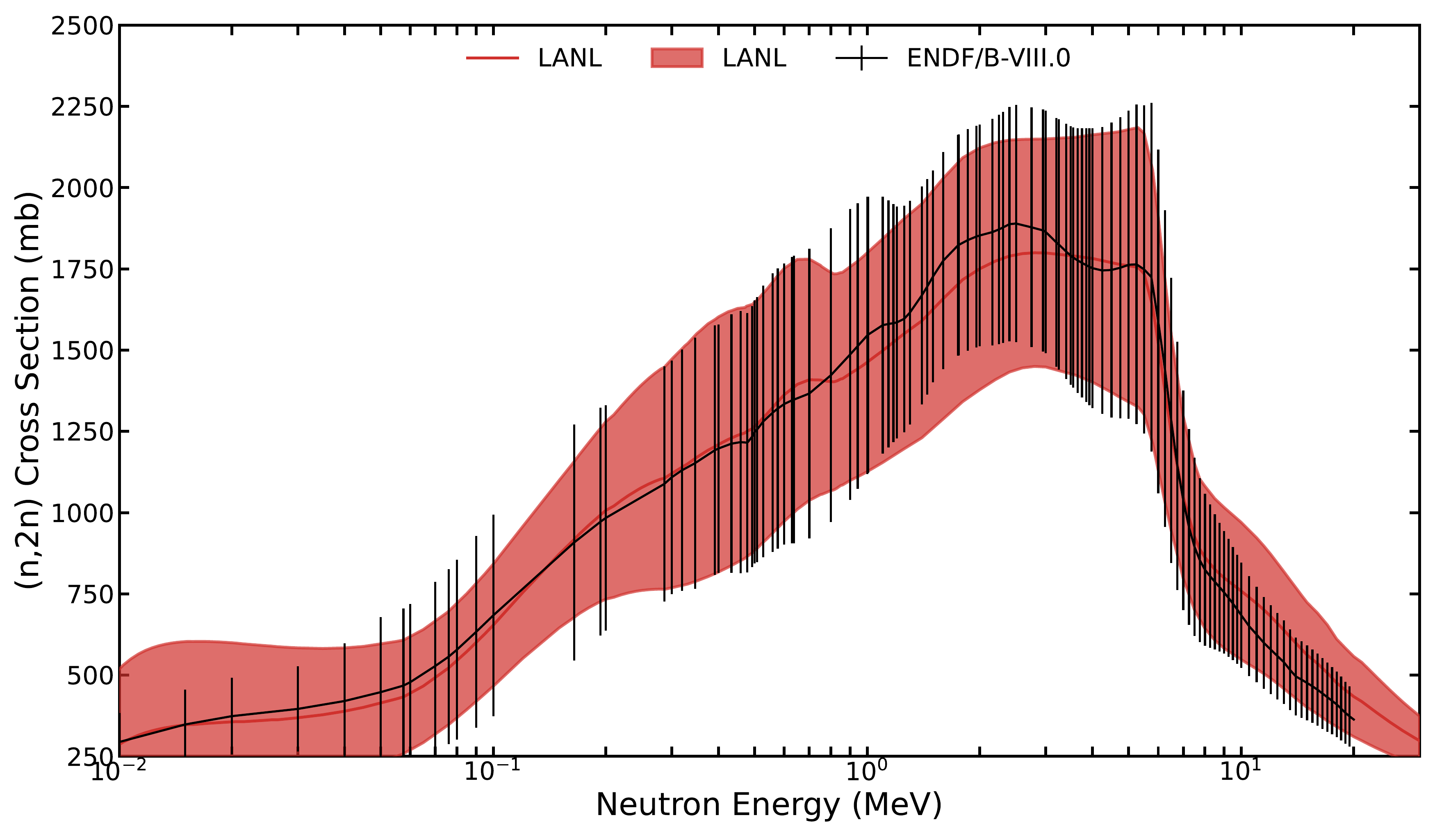}
\caption{Variance for inelastic cross section.}
\label{fig:cov_239pu_cs_inel}
\end{figure}

\begin{figure}
\centering
\includegraphics[width=0.49\textwidth]{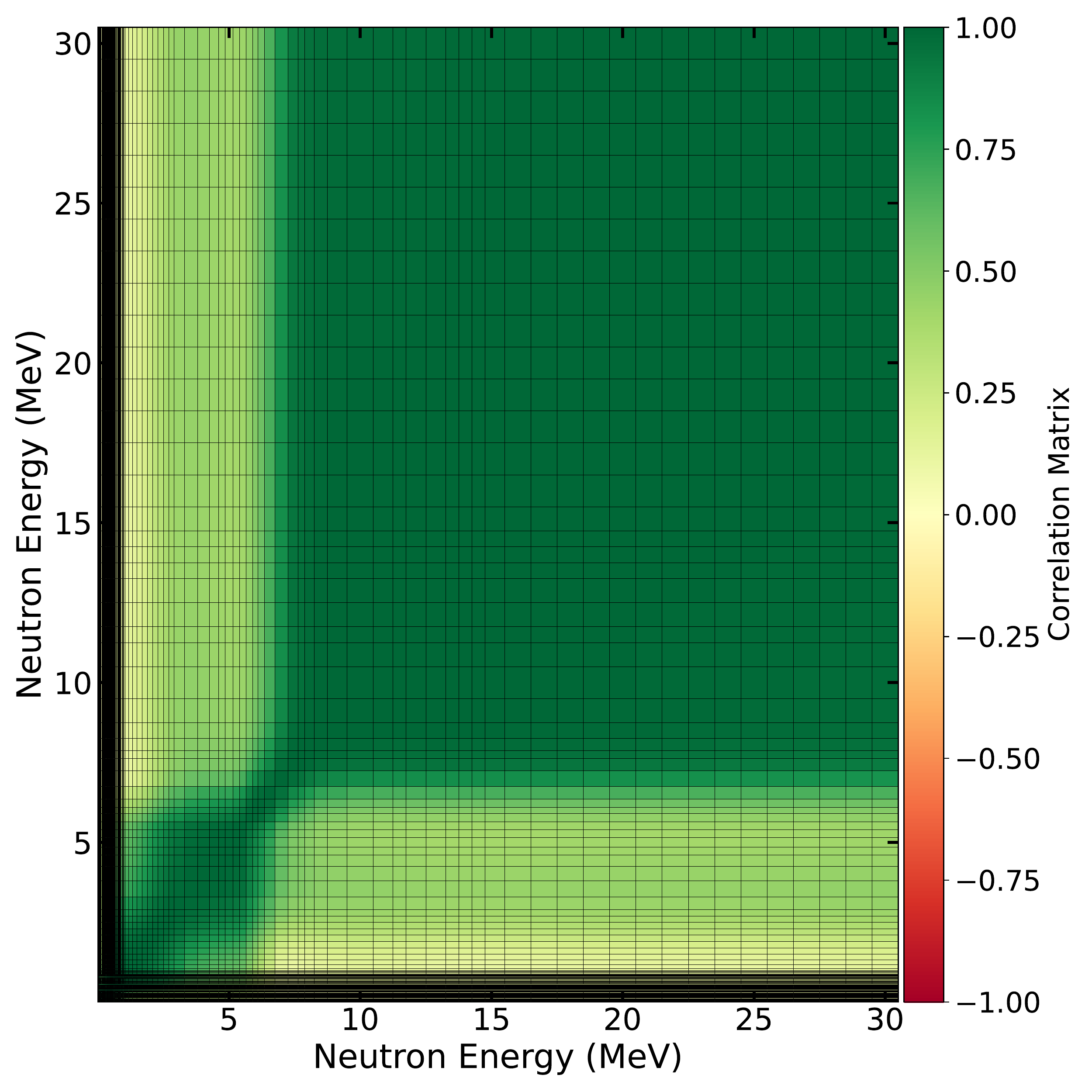}
\includegraphics[width=0.49\textwidth]{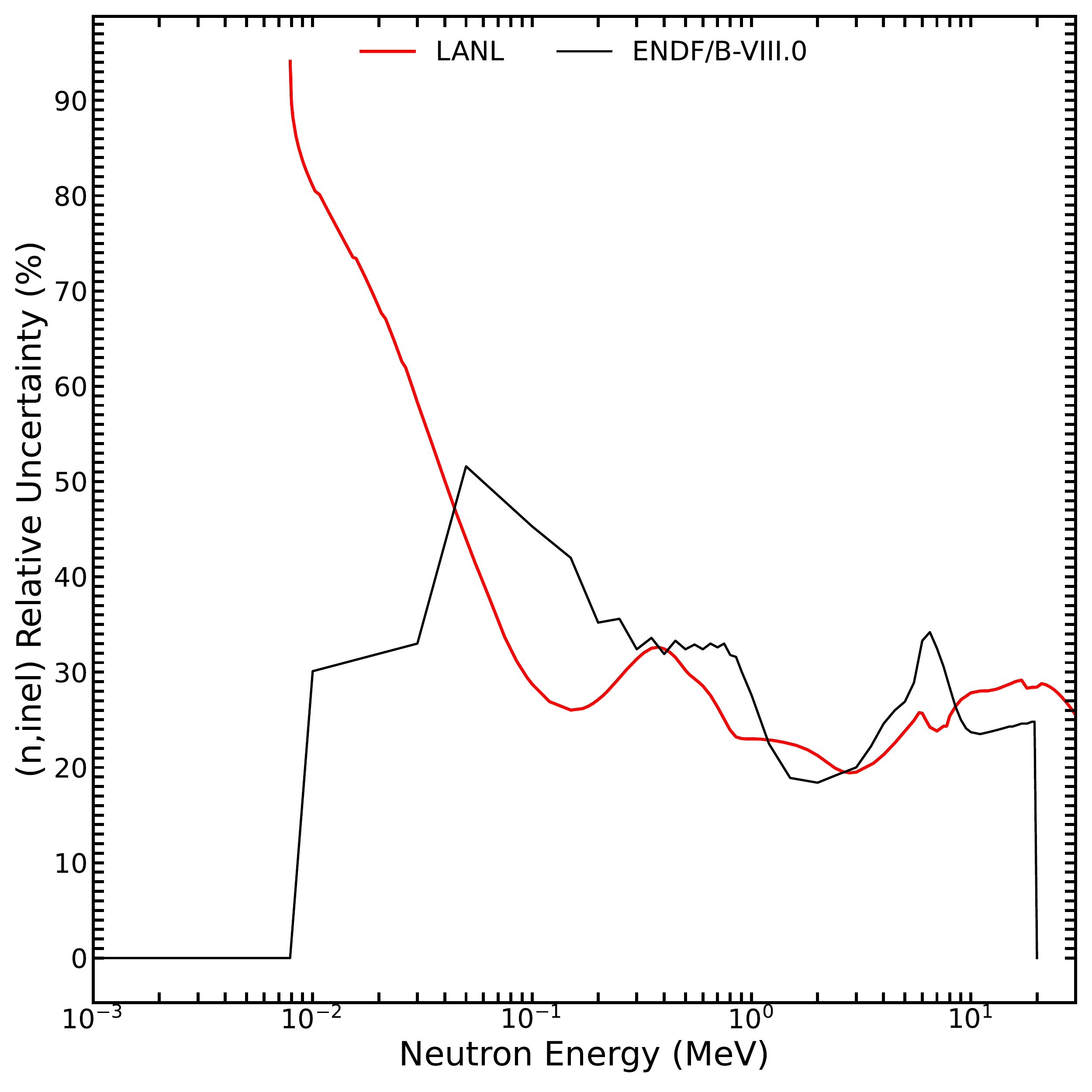}
\caption{(a) Correlation matrix for inelastic cross section. (b) Relative uncertainty as a function of incident energy.}
\label{fig:cov_239pu_inel}
\end{figure}

Figure \ref{fig:cov_239pu_cs_inel} shows the variances (square root of the diagonal entries of the covariance matrix) along the evaluated inelastic cross section. 
Figure \ref{fig:cov_239pu_inel} (a) shows the correlation matrix for the evaluated inelastic cross section and (b) shows the relative uncertainties as a function of incident neutron energy for the evaluated inelastic cross section. 

\begin{figure}
\centering
\includegraphics[width=0.9\textwidth]{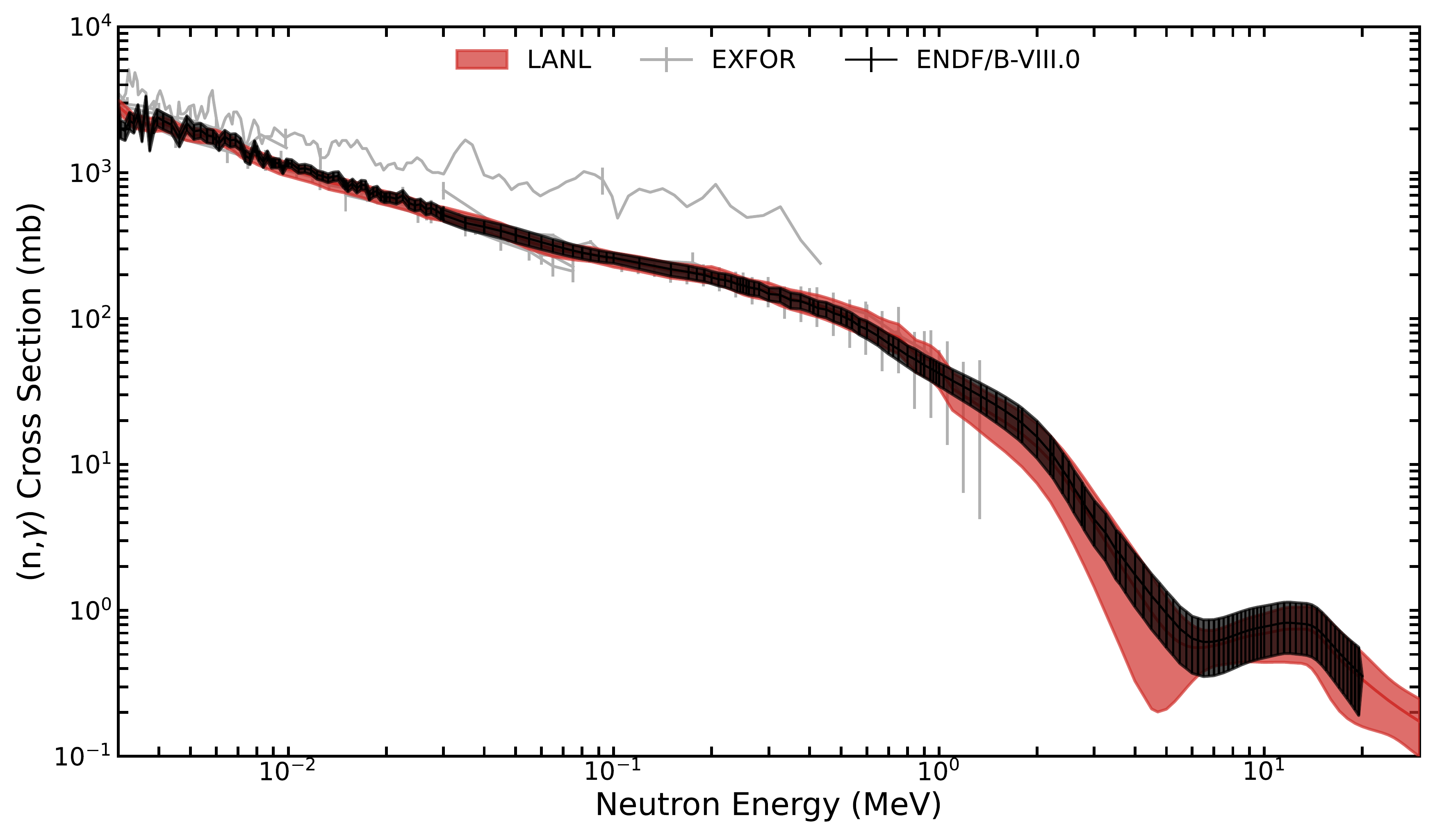}
\caption{Variance for the (n,$\gamma$) cross section.}
\label{fig:cov_239pu_cs_ng}
\end{figure}

\begin{figure}
\centering
\includegraphics[width=0.49\textwidth]{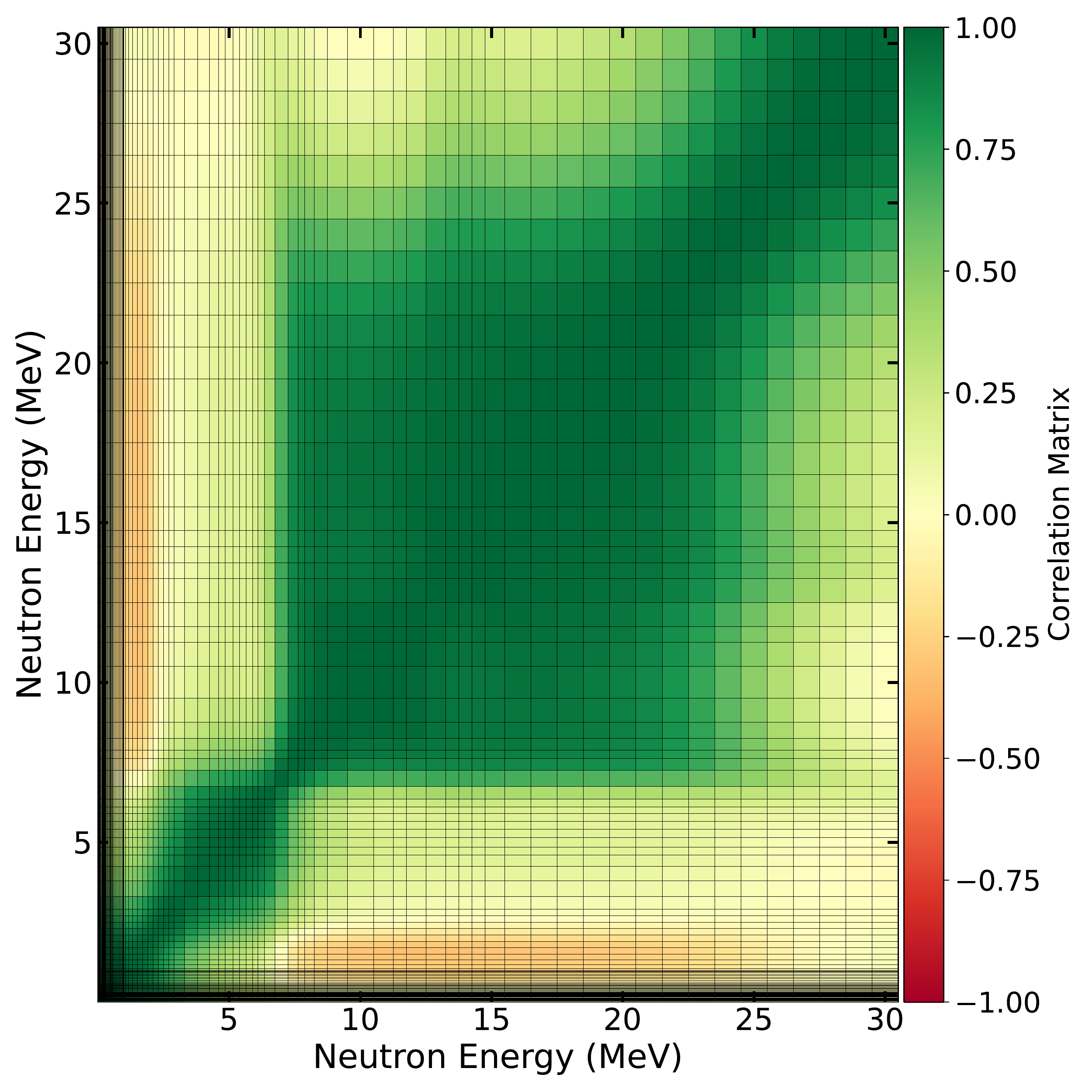}
\includegraphics[width=0.49\textwidth]{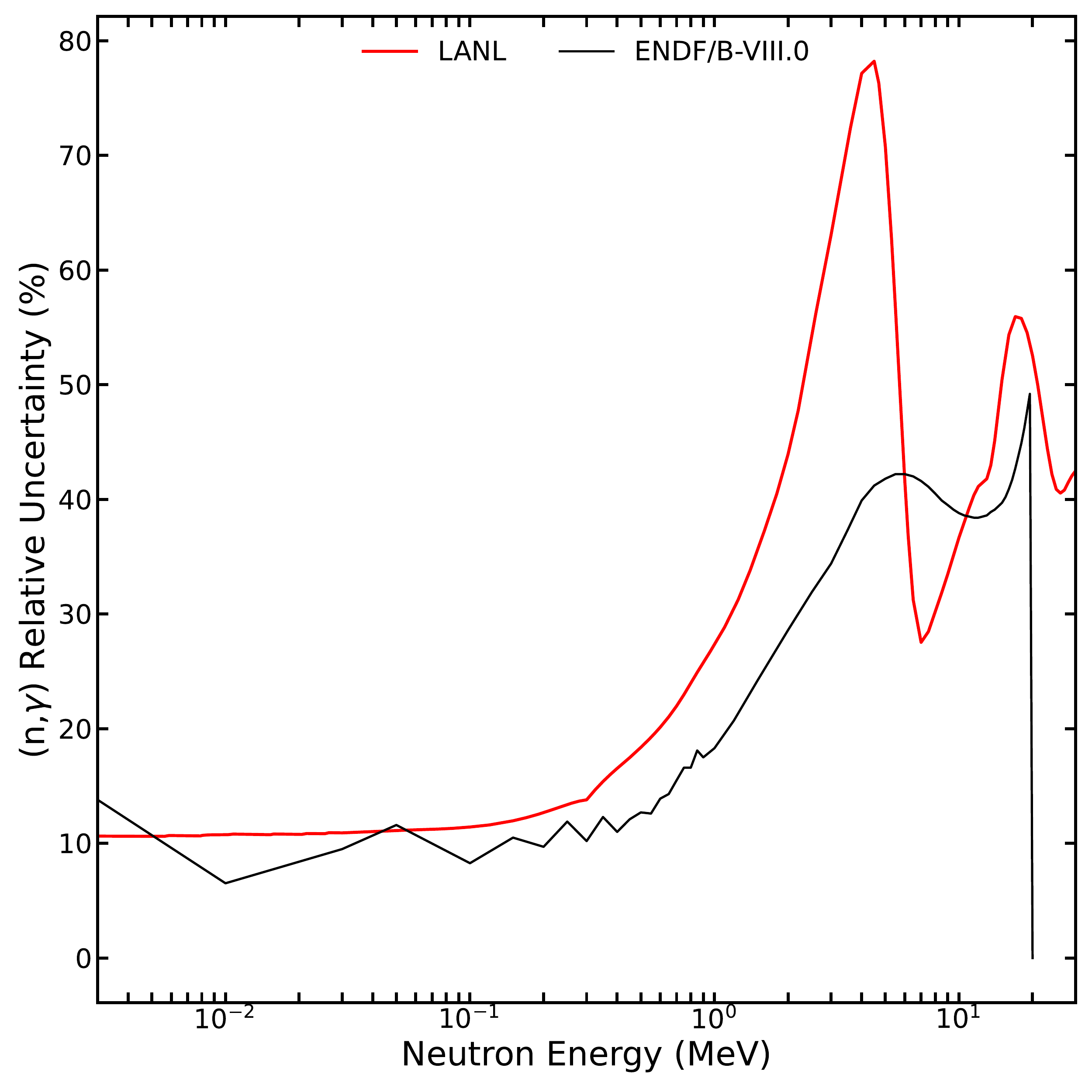}
\caption{(a) Correlation matrix for (n,$\gamma$) cross section. (b) Relative uncertainty as a function of incident energy.}
\label{fig:cov_239pu_ng}
\end{figure}

Figure \ref{fig:cov_239pu_cs_ng} shows the variances (square root of the diagonal entries of the covariance matrix) along the evaluated (n,$\gamma$) cross section. 
Figure \ref{fig:cov_239pu_ng} (a) shows the correlation matrix for the evaluated (n,$\gamma$) cross section and (b) shows the relative uncertainties as a function of incident neutron energy for the evaluated (n,$\gamma$) cross section. 

\begin{figure}
\centering
\includegraphics[width=0.9\textwidth]{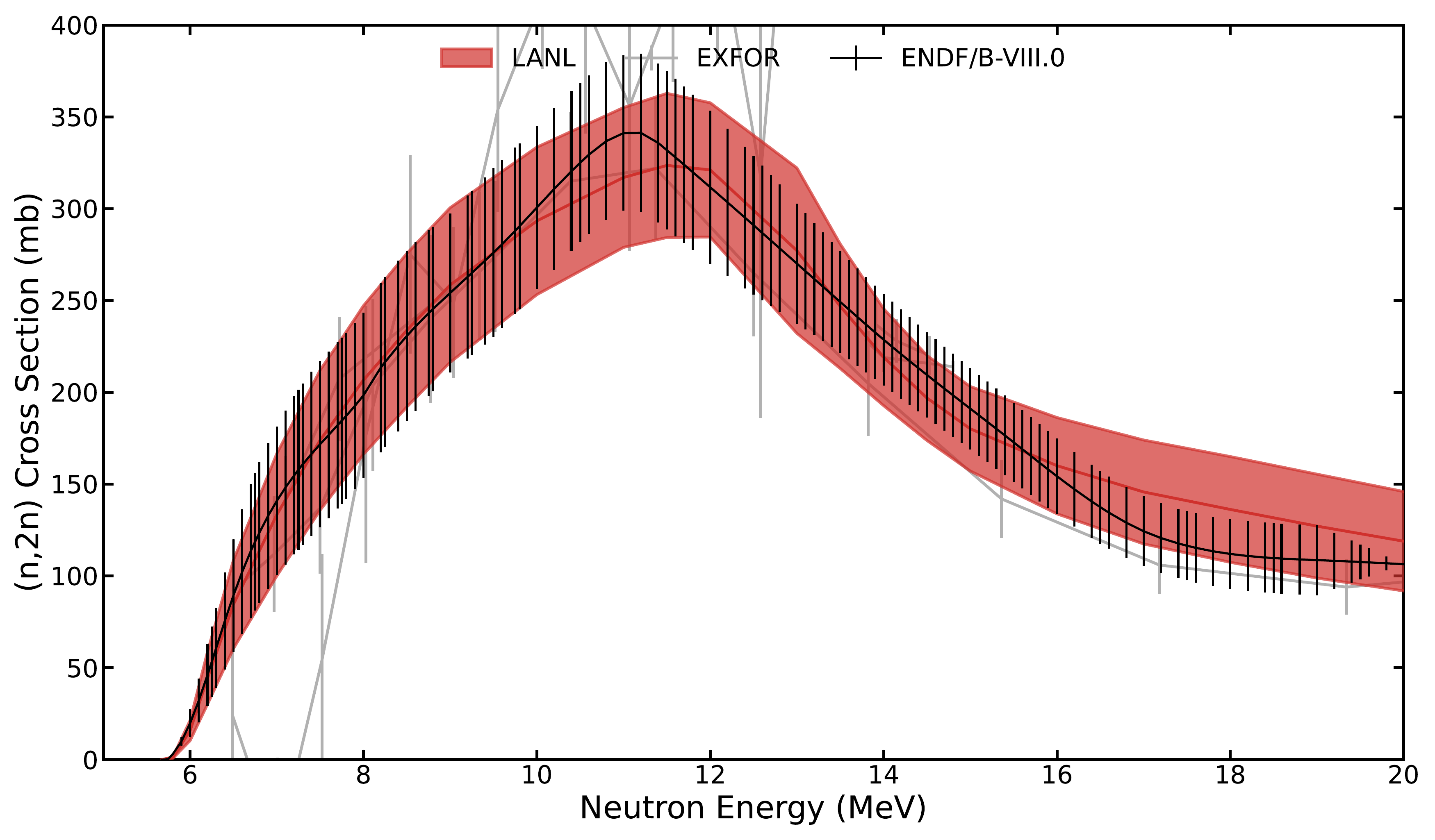}
\caption{Variance for the (n,2n) cross section.}
\label{fig:cov_239pu_cs_n2n}
\end{figure}

\begin{figure}
\centering
\includegraphics[width=0.49\textwidth]{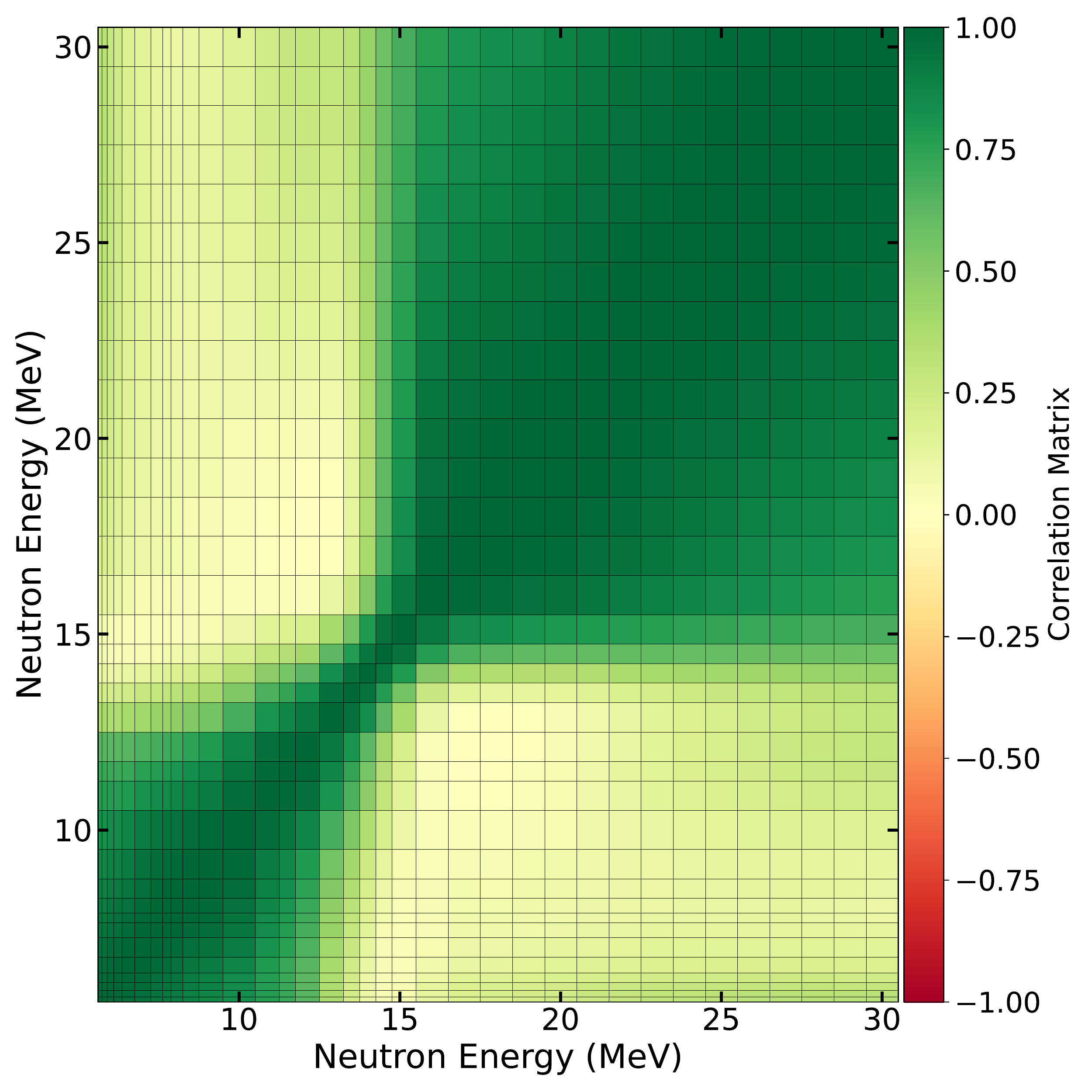}
\includegraphics[width=0.49\textwidth]{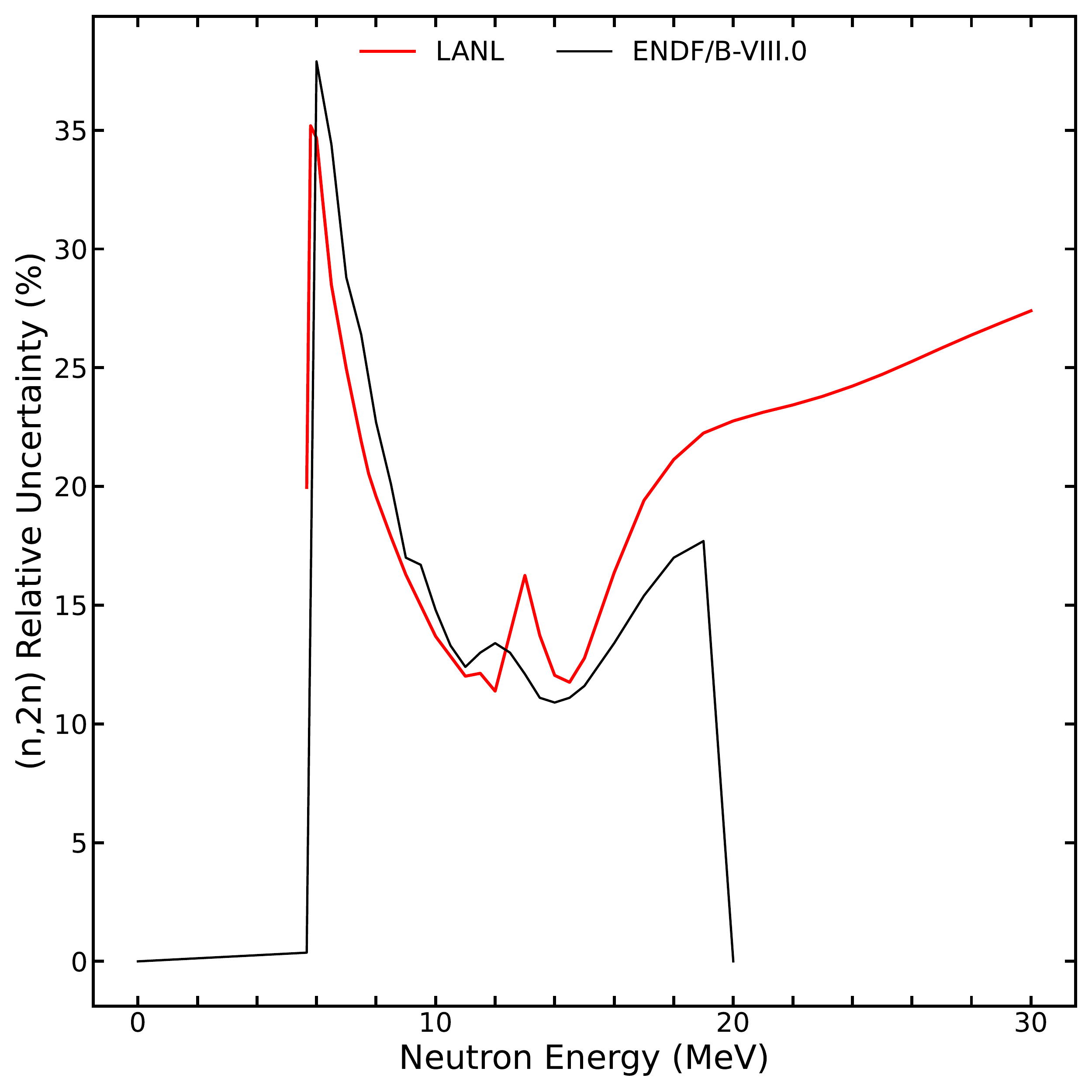}
\caption{(a) Correlation matrix for (n,2n) cross section. (b) Relative uncertainty as a function of incident energy.}
\label{fig:cov_239pu_n2n}
\end{figure}

Figure \ref{fig:cov_239pu_cs_n2n} shows the variances (square root of the diagonal entries of the covariance matrix) along the evaluated (n,2n) cross section. 
Figure \ref{fig:cov_239pu_n2n} (a) shows the correlation matrix for the evaluated (n,2n) cross section and (b) shows the relative uncertainties as a function of incident neutron energy for the evaluated (n,2n) cross section. 
Note the slight decrease to the relative uncertainties due to the inclusion of the new Meot data. 

\section{Points of contact}

MRM is responsible for the evaluation methodology. 
TK is the point of contact for the Los Alamos statistical Hauser-Feshbach code, \CoH{}. 
DN and NK performed the validation and testing. 
DN evaluated the (n,f) cross section, and PFNS included in the file that was validated, while DN, AEL and PT evaluated the $\overline\nu_p$. 
AEL and IS assisted with covariance and uncertainty methodologies. 

\section{Acknowledgments}
MM thanks R.~Capote for valuable discussions. 
MM thanks M.~B.~Chadwick for emphasizing the importance of measured data. 
Work at LANL was carried out under the auspices of the National Nuclear Security Administration (NNSA) of the U.S. Department of Energy (DOE) under contract 89233218CNA000001. 

\bibliographystyle{unsrt}
\bibliography{refs}

\end{document}